\documentclass[a4paper,11pt]{article}
\usepackage[title]{appendix}

\usepackage[utf8]{inputenc}
\usepackage[affil-sl,auth-sc]{authblk}
\usepackage{amssymb,amsmath,amsbsy}
\usepackage{bm}
\usepackage{appendix}
\usepackage[linktocpage=true]{hyperref}
\usepackage[top=1.3in, bottom=1.3in,left=0.9in,right=0.9in]{geometry}
\usepackage{graphicx}
\usepackage{cite}
\usepackage{float}
\usepackage{caption}
\usepackage{wasysym}
\usepackage{float}
\usepackage{slashed}
\usepackage{caption}
\usepackage{indentfirst}
\usepackage{multirow}
\usepackage{feynmp-auto}
\usepackage{braket}
\usepackage[nottoc]{tocbibind} 
\usepackage{color}
\usepackage{array}
\usepackage{epstopdf} 

\let\originalleft\left \let\originalright\right
\renewcommand{\left}{\mathopen{}\mathclose\bgroup\originalleft}
\renewcommand{\right}{\aftergroup\egroup\originalright}

\numberwithin{equation}{section}

\begin{document}

\title{\bf SM EFT effects in Vector-Boson Scattering \\ at the LHC}

 \author{A. Dedes$^1$, P. Koz{\'o}w$^2$, M. Szleper$^3$}

\affil{\small $^1$ Department of Physics, Division of Theoretical
  Physics, \\ University of Ioannina, GR 45110, Greece}
  
\affil{\small $^2$ Institute of Theoretical Physics, Faculty of Physics, University of Warsaw, 
\\  ul.~Pasteura 5, 02--093 Warsaw, Poland}

\affil{\small $^3$ National Center for Nuclear Research, High Energy Physics Department,\\
 ul.~Pasteura 7, 02-093 Warsaw, Poland }

\date{}

\maketitle
\thispagestyle{empty}

\begin{abstract}
 The assumption that the Standard Model  is an Effective Field Theory (SM EFT)
of a more fundamental theory  at a higher, than electroweak, energy scale, 
implies  a growth of cross-sections for electroweak Vector Boson Scattering (VBS) processes  signalling the appearance of a resonance
(or resonances) nearby that scale.
In this article, we investigate in detail SM EFT effects 
from dimension-6 operators in VBS with  like-sign-$W$ production in 
 fully leptonic decay modes at the High Luminosity LHC (HL-LHC).
We find that these effects are  important for a handful of operators, most notably
 for the operator composed of three $SU(2)$ field strength tensors responsible
for strong transversely polarized vector boson interactions.
Current global fits on Wilson-coefficients allow for a signal immediately
 permissive at the HL-LHC if not accessible at the current LHC-dataset.

\end{abstract}


\section{Introduction}
\label{sec:introduction}
It is widely believed that the Standard Model (SM)~\cite{Weinberg:1967tq,Glashow,Salam} is an Effective, low energy, Field Theory (EFT) approximation~\cite{Weinberg:1980wa,Callan:1969sn,Coleman:1969sm} to a more fundamental theory (UV-theory) which overtakes the SM
at energies, at least several times, higher than the Electroweak (EW) scale.
One of the very important processes that may shed light on the dynamics of the UV-theory is the  electroweak Vector Bosons Scattering  (VBS)\footnote{For a review see~\cite{Green:2016trm}.}. This claim
is true because,
\begin{itemize}
	\item[(i)] VBS processes
are directly related to the mechanism of electroweak
symmetry breaking, and as such,  regarded as complementary to Higgs-boson measurements at LHC,

\item[(ii)] although inaccessible directly,  heavy particles, constituents of the UV-theory, may leave their trace in modifying the strength and 
dynamics  of electroweak interactions, 
spoiling for example the, at most constant, center of mass energy ($s$)-behaviour of the SM $VV\rightarrow VV$ ($V=W^\pm, Z,\gamma$) amplitudes. 
In the SM EFT,  this results in a growth of cross-sections at energies straight after the electroweak scale.

\item[(iii)] In the SM, even when next-to-leading order corrections are included, VBS processes feature particularly slow slope with energy in comparison to other electroweak processes~\cite{Biedermann:2016yds}.
\end{itemize}
Therefore, points (ii) and (iii) imply potential sensitivity to the UV-theory particles,
 especially to those strongly interacting with  EW-gauge bosons.

Unfortunately, there is no $VV$-collider available\footnote{Interestingly in a certain setup, it has been argued that a muon collider could effectively be regarded as such~\cite{Buttazzo:2018qqp,Costantini:2020stv}.}, however the $VV\rightarrow VV$ reactions are indirectly accessible at the LHC, particularly within its high luminosity phase (HL-LHC), through the process $pp \rightarrow 2\text{jets} + 2\text{lepton pairs}$.
\begin{figure}
\begin{center}
\includegraphics[scale=0.5]{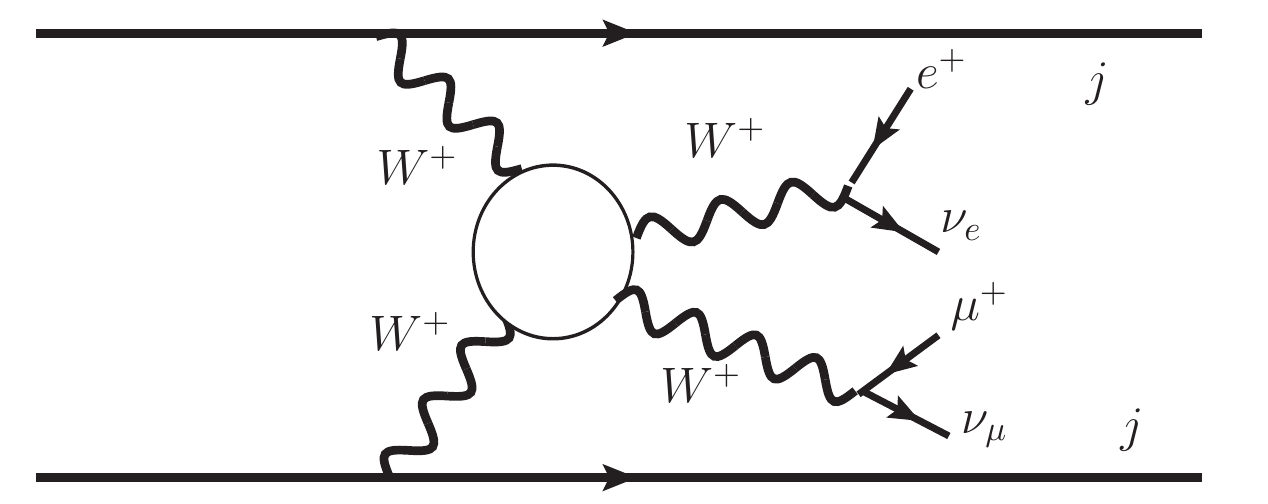} 
\caption{\sl A Feynman diagram picturing a VBS process.} 
\label{fig:VBS}
\end{center}
\end{figure}
While all the elastic processes e.g. $ZZ,WZ, W^\pm W^\pm, W^\pm W^\mp$  have been extensively investigated at the LHC, the  same-sign $W^+W^+$ process in fully-leptonic decay mode 
has been the first process to be observed\footnote{The actual measurement involves a sum of both $W^+W^+$ and $W^-W^-$ processes; $W^+W^+$ 
makes about 4/5 of it~\cite{Sirunyan:2020gyx}.} 
at $5\sigma$~\cite{Sirunyan:2017ret} (based on the 13 TeV dataset), 
and is currently confirmed at a sensitivity far above $5 \sigma$~\cite{Sirunyan:2020gyx}. For that reason, in this paper we shall investigate in detail the VBS process through 
$W^+W^+$-scattering, namely 
the reaction $$pp \ \rightarrow \  2\,\text{jets} + W^{+\ast}\, W^{+\ast} \ \rightarrow \ 
2\,\text{jets} + 2\,\text{charged-leptons} + 2\,\text{neutrinos}\;.$$ 
 This is exemplified by Fig.~\ref{fig:VBS} of a particular final state. 
Throughout this work, we mostly focus on  the HL-LHC experimental perspectives.

We follow the SM EFT approach\footnote{For a review see ref.~\cite{Brivio:2017vri}.} in which the corrections from heavy and decoupled UV-states~\cite{Appelquist:1974tg},  can be parametrized
by   a number of $i$-indexed dimension-6 operators, $Q_i$, added to the SM Lagrangian, associated with dimensionless (or dimensionful)
Wilson coefficients $C^i$  (or $f^i$), as
\begin{equation}
{\cal L} \ = \   {\cal L}_{SM}+\sum_i\,\frac{C^i}{\Lambda^2} \, Q_i + \dots \
 \equiv \  {\cal L}_{SM} + \sum_i f^i \, Q_i + \dots  \;.
\label{eq:lagrangian}
\end{equation} 
The mass scale $\Lambda$ denotes the lightest mass among the heavy 
particle masses within the UV-theory. Although the EFT expansion in \eqref{eq:lagrangian} is written up-to $1/\Lambda^2$, in one occasion below we will fill in the $``\dots "$ with several dimension-8 operators where the dimensionless Wilson coefficients are suppressed by four powers of $\Lambda$'s.

The real advantage of the EFT approach follows from the fact that one can study the discovery potential
of the physics Beyond the SM (BSM) without any knowledge of  theories that may lay ahead by making use of a limited number of  operators arranged order-by-order in  $1/\Lambda$ expansion. More precisely, here we consider the 
SM EFT theory that matches UV-models featuring linearly realized electroweak symmetry breaking, where the Higgs field is part of the SM Higgs doublet. 
The SM EFT basis of non-redundant operators has been constructed first at dimension-6, commonly referred to as Warsaw basis~\cite{Grzadkowski:2010es}, and very recently at  dimension-8~\cite{Murphy:2020rsh,Li:2020gnx}. 

It is a common approach, in both experimental and theoretical literature, to use EFT for studying VBS channels as an indirect search for physics Beyond the Standard Model (BSM)~\cite{Bagger:1993zf,Espriu:2012ih,Chang:2013aya,Kilian:2014zja,Brass:2018hfw}; for a recent review see~\cite{Baglio:2020bnc}. The point of
focus of such studies is usually the $VVVV$ quartic-couplings, i.e. the operators that modify Quartic Gauge Couplings (QGC) and simultaneously leave
intact Trilinear Gauge Couplings (TGC) and Higgs-gauge boson interactions.   
In  SM EFT such kind of physics arise from dimension-8 operators~\cite{Degrande:2013rea,Eboli:2016kko} and, consequently, the searches are typically conducted as if there was no effect arising from dimension-6 operators. Here, we investigate the validity of this assumption, given experimental constraints on dimension-6 interactions (which necessarily modify TGC) emerging from independent channels, by studying in detail 
the like-sign $W$-boson production through VBS at (HL-)LHC. We would like to emphasise that the aim of this work is not to model accurately possible BSM signals, but to check which dimension-6 operators can produce non-negligible effect as allowed by the most up to date experimental constraints from non-VBS processes on the corresponding Wilson coefficients (including also the important issue of bounding background operators), which include new LHC Run 2 results.  
The numerical significance of  dimension-6 operators in VBS has been pointed out  
 in  Ref.~\cite{Fabbrichesi:2015hsa,Gomez-Ambrosio:2018pnl}.

Concerning the bounds on dimension-6 operators, 
we use the ones reported in Refs.~\cite{Dawson:2020oco} and~\cite{Ellis:2018gqa} (non-4-fermion operators) and~\cite{Domenech:2012ai,Sirunyan:2017ygf}(4-fermion operators)\footnote{Notice, however, the discussion in ref.~\cite{Alte:2017pme}}.
In~\cite{Dawson:2020oco,Ellis:2018gqa}, the truncation of EFT cross-sections is performed consistently at dimension-6, namely, no (dimension-6)$^2$ terms are considered, that would arise from squaring the $(\text{SM}+\text{dimension-6})$ amplitude. It is known that inclusion of the latter significantly improves the constraints from di-boson production channels. However, dimension-8 effects in the EFT expansion are then generally expected to be significant as well, if the underlying UV interactions are not particularly strong. Therefore, inclusion of (dimension-6)$^2$ terms in principle implies losing model-independence within EFT~\cite{Falkowski:2016cxu}. On the other hand, there is certain sensitivity to the dimension-6 Wilson coefficients in the cross-sections and the reported limits are, therefore, to be understood as conservative constraints. 

There are many LHC analyses, a partial list includes 
Refs.~\cite{Aaboud:2019nmv,Sirunyan:2017ret,Aaboud:2018ddq,Sirunyan:2020gyx}. Thus far, many $VV\to VV$ processes have been discovered, but within errors agree
with the SM.  Interestingly,  current experimental precision does not constrain the ``weak coupling'' regime, leaving plenty of space for new physics effects. It is particularly attractive since the uncertainties after LHC Run 2 are (by far) ``statistics-dominated''~\cite{Sirunyan:2020gyx}.

The paper is organized as follows. In section~\ref{sec:W}, we present the 
relevant EFT-operators and motivate theoretically, by using analytical formulae from 
Appendix~A, the more technical subsequent analysis. The core numerical 
analysis for VBS at (HL-)LHC is presented in section~\ref{sec:anal}. We
conclude with section~\ref{sec:conclusions}.

\section{Warming-up: $W^+W^+ \to W^+ W^+$ scattering}
\label{sec:W}

\begin{figure}
\begin{center}
\includegraphics[scale=0.5]{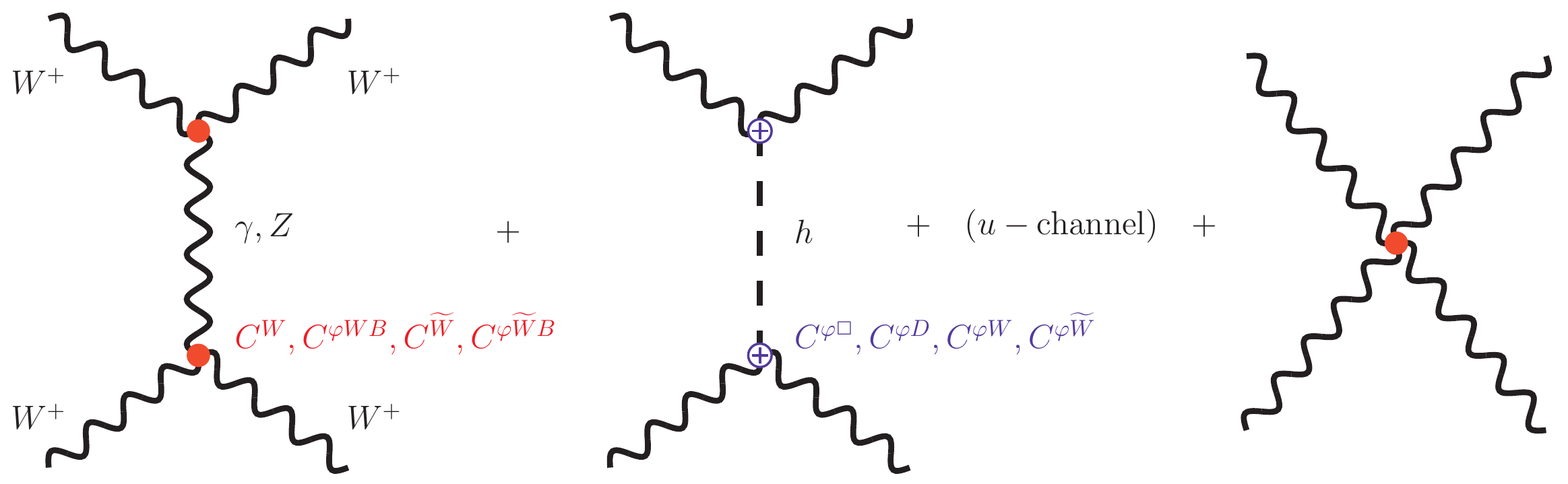} 
\caption{\sl Tree level Feynman diagrams filling in the white blob of Fig.~\ref{fig:VBS}. Associated Wilson-coefficients to operators affecting each vertex are also shown.}
\label{fig:WWdiags}
\end{center}
\end{figure}

As we mention in the introduction, although LHC experimental analyses of VBS are optimized for QGCs (dimension-8 operators), 
it is important, if not necessary, to examine the impact on TGCs (dimension-6 operators) on ``golden'' process  $W^+W^+\to W^+W^+$.\footnote{In section~\ref{sec:anal} we
also examine the effect of ``background'' dimension-6 operators entering in other sub-processes.} 
After all, dimension-6 operators arise at leading order in EFT expansion.
At tree level in SM EFT, as we prove explicitly, the process $W^+W^+\to W^+W^+$  is gauge invariant,  that is independent of the gauge fixing parameter, 
and hence its self-study is at least legitimate. 
In both SM and SM EFT at leading order there are seven Feynman diagrams
  shown in Fig.~\ref{fig:WWdiags},
mediated by a photon, a $Z$ and a Higgs-boson in both $t$- and $u$-channels plus a contact diagram. 
Following closely the Warsaw basis notation~\cite{Grzadkowski:2010es}, 
the Lagrangian~\eqref{eq:lagrangian} 
is affected at tree-level  by 5 CP-Conserving and 3 CP-Violating, dimension-6, operators arranged in Table~\ref{tab:ops}.
\begin{table}
\begin{center}
\begin{tabular}{|c|c|c|c|}
\hline
  & $X^3$   & $\varphi^4 D^2$ & $X^2 \varphi^2$\\[1mm] \hline 
CPC & 
$Q_W =\epsilon^{IJK} W_\mu^{\nu\, I} W_\nu^{\rho \, J} W_\rho^{\mu\, K}$ &   
$Q_{\varphi\Box} = (\varphi^\dagger \varphi) \Box (\varphi^\dagger \varphi)$ &
$Q_{\varphi W} = \varphi^\dagger \varphi \, W_{\mu\nu}^I W^{\mu\nu\, I}$ \\ 
  &   & $Q_{\varphi D} = (\varphi^\dagger D^\mu \varphi)^*\, (\varphi^\dagger D_\mu \varphi)$ & $Q_{\varphi WB} = \varphi^\dagger \tau^I \varphi \, W_{\mu\nu}^I B^{\mu\nu}$ \\[2mm] \hline
CPV &
$Q_{\widetilde{W}} =\epsilon^{IJK} \widetilde{W}_\mu^{\nu\, I} W_\nu^{\rho \, J} W_\rho^{\mu\, K}$ &   &  $Q_{\varphi \widetilde{W}} = 
\varphi^\dagger \varphi\, \widetilde{W}_{\mu\nu}^I W^{\mu\nu\, I}$ \\
 & & & $Q_{\varphi \widetilde{W} B} = 
\varphi^\dagger  \tau^I \varphi\, \widetilde{W}_{\mu\nu}^I W^{\mu\nu\, I}$ \\ \hline
\end{tabular}
\caption{\sl Dimension-6 operators, in the Warsaw basis, modifying the process $W^+W^+ \to W^+W^+$.}
\label{tab:ops}
\end{center}
\end{table}

The  leading high energy helicity amplitudes are given in Appendix~\ref{app:A}. We work in the Warsaw basis and 
use the Feynman rules in $R_\xi$-gauges from ref.~\cite{dedes:2017zog}.
 The operators $Q_{\varphi W B},$ and $Q_{\varphi \widetilde{W} B}$ disappear  from leading-$s$  helicity amplitudes in $WW\to WW$ processes. 
This is easily seen by just looking at the SM EFT Feynman Rules and use  the Goldstone 
boson equivalence theorem~\cite{Cornwall:1974km,Vayonakis:1976vz,Lee:1977eg,Chanowitz:1985hj,Gounaris:1986cr,Yao:1988aj,Bagger:1989fc}, where
for these operators and specifically to $WW \to WW$ channel, there are no associated contact four-point interactions 
involving Goldstone bosons nor exist a contact interaction with four $W$'s proportional to $C^{\varphi WB}$ or $C^{\varphi \widetilde{W} B}$.
One may think that there may be contributions from gluing the 3-point vector boson vertices shown in Fig.~\ref{fig:WWdiags} where these insertions exist,
but it can be proved explicitly that any $s$-enhanced amplitude cancels out when consistently expanding the $Z$-mass, also affected by $C_{\varphi WB}$.
%
Another feature of importance  is
  the appearance of the $t$-channel enhancement 
   $(1-\cos^2\theta)$, where $\theta$ is the scattering angle, in the denominator of
 leading SM amplitudes, see, for example $\mathcal{M}_{++++}$ in \eqref{eq:M++++}.
   This is in contrast to the fact that none of SM EFT amplitudes has such 
   a $t$-channel enhanced factor   that is accompanied by a growth of energy, 
   see  eqs.~\eqref{amp:ppppTTTT}-\eqref{amp:ppppCPV}.

Furthermore, 
in Appendix~\ref{app:1}  we also arrange analytical expressions for the helicity 
cross sections following the notation where 
 $``T"$ stands for transverse helicities $\pm 1$ and $``L"$ for longitudinal 
gauge bosons with helicity-0. 
In the SM, the dominant polarized cross-sections come into the following rates: 
\begin{equation}
\sigma_{TTTT} : \sigma_{LLLL} : \sigma_{LTLT} : \sigma_{TLTL} : \sigma_{TLLT} :
\sigma_{LTTL}
\ \approx \ 1 : \frac{1}{8.5} : \frac{1}{8.0} : \frac{1}{8.0} : \frac{1}{8.0} : \frac{1}{8.0} \;.
\end{equation}
As we already mentioned, in the SM all polarized 
cross sections are enhanced by the $t$-channel  factor 
but, particularly for the $TTTT$-mode there is an accidentally enhanced factor of  8 
w.r.t. the other modes as can easily be seen from eqs.~\eqref{eq:WpTTTT}-\eqref{eq:WpLTLT}.

 In SM EFT there are interference effects between the SM and dimension-6 operators
only in the $LLLL$-mode (or ``0000'' mode). CP-violating contributions  enter in the mixed and pure transverse channels, similar to their CP-conserving counterparts. 
The cross sections have the
 symbolic form $\sigma \sim \mathrm{SM}^2 + \mathrm{SM}\times \mathrm{dim6} + \mathrm{dim6}^2$. Following this pattern we obtain ($\bar{g}^2$ is the $SU(2)_L$ gauge coupling):
 \begin{align}
 \sigma_{TTTT}(s) & \approx  \frac{\bar{g}^4}{s} \biggl [ \frac{A_{T}}{1-c^2} \ + \
   B_{T} \cdot 0 \ + \ \Gamma_{T} \: \bar{g}^2 \: \left ( \frac{|C^W|}{\bar{g}^2} \right )^2\:
 \left ( \frac{s}{\Lambda^2} \right )^2 \ + \ \cdots \biggr ]\;, \label{eq:WpsTTTT}
  \\[2mm]
  \sigma_{LLLL}(s) & \approx  \frac{\bar{g}^4}{s} \biggl [ \frac{A_{L}}{1-c^2} \ + \
   B_{L} \: \left ( \frac{C^{\varphi\Box}}{\bar{g}^2} \right )\: \left ( \frac{s}{\Lambda^2} \right )   
   \ + \ \Gamma_{L} \: \left (\frac{C^{\varphi\Box}}{\bar{g}^2} \right )^2\:
 \left ( \frac{s}{\Lambda^2} \right )^2 \ + \ \cdots \biggr ]\;, \label{eq:WpsLLLL}
 \end{align}
where $A_i,B_i,\Gamma_i\, (i=T,L)$ are dimensionless coefficients read from eqs.~\eqref{eq:WpTTTT} and \eqref{eq:WpLLLL} 
 that depend upon the cutting angle $c\equiv \cos(\theta_{cut})$  and  ratios of
vector boson and Higgs masses,  and ``$\dots$'' are corrections from higher than six dimensional operators. The corresponding expression for $\sigma_{LTLT}$ and related cross-sections are similar in form with
 \eqref{eq:WpsTTTT}. 
In order to justify clearly our points, 
we focus only on $C^W$ and $C^{\varphi\Box}$  contributions 
in eqs.~\eqref{eq:WpsTTTT}-\eqref{eq:WpsLLLL}. Full analytical 
expressions are given in Appendix A,  eqs.~\eqref{eq:WpTTTT}-\eqref{eq:WpLTLT}.

In  a \textit{weakly coupled} UV-theory with perturbative decoupling,  dimensional analysis~\cite{Arzt:1994gp} results in $C^W \approx g^3/(4\pi)^2, C^{\varphi \Box} \approx g^2$. These effects could be important only  if $A_L \approx (1-c^2) B_L (s/\Lambda^2)$ which is never the case for $c\approx 1$ 
and $s<\Lambda^2$, which suggests weak sensitivity to small coupling regime
 in the  $W^+W^+$ process. This result, however, 
does strongly depend on the cutting-angle $\theta_{cut}$. 
For example, in differential cross-section 
distributions for the $LLLL$-mode, there are zeros 
for different values of SM EFT $C^{\varphi \Box}$ and/or $C^{\varphi D}$ input values!
This case study is  however statistics limited at LHC mainly because
 the SM dominant $TTTT$-mode will not be  affected, but may be  
important for future  HL-LHC studies.
%

%
\begin{figure}
\begin{center}
\mbox{
\includegraphics[scale=0.6]{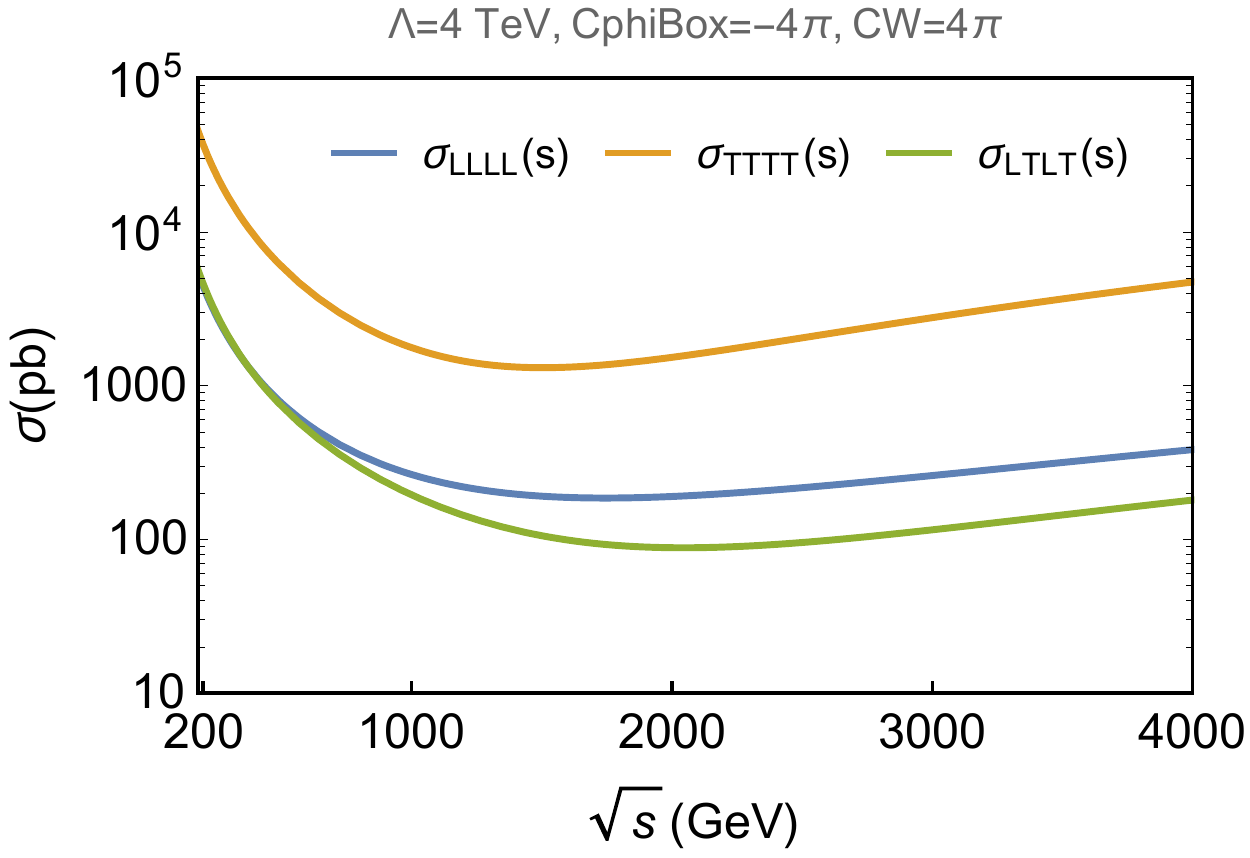} 
\includegraphics[scale=0.6]{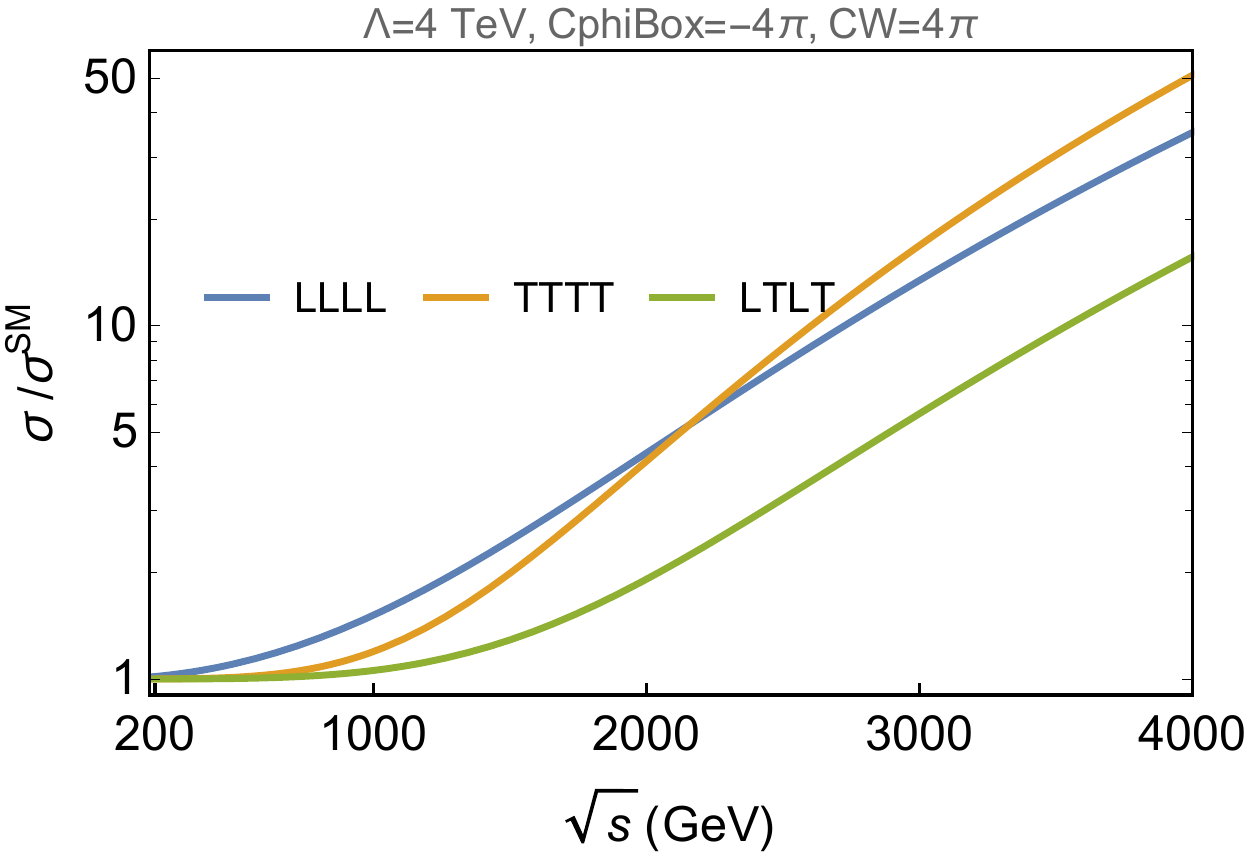} 
}
\end{center}
\caption{\sl Left: Polarized cross sections for the three mode cross sections,
 eqs.~\eqref{eq:WpTTTT}-\eqref{eq:WpLTLT}, as a function of the 
c.m energy $\sqrt{s}$. We have chosen as input values the coefficients shown at
the top of the figures and $\theta_{cut}=\pi/18$. Right: Similarly for the ratios of the 
polarized cross sections w.r.t  the SM result.}
\label{fig:Wp}
\end{figure}

In a \textit{strongly coupled} UV-theory or in a UV-theory with \textit{composite gauge bosons}
the loop-suppression of $C^W$ may not be dictated by gauge invariance.
Naive dimensional analysis~\cite{Gavela:2016bzc} in such 
an extreme case may result in SM EFT coefficients as big as 
$C^W \approx 4\pi$ and $C^{\varphi\Box} \approx (4\pi)^2$.
What is most important here is the fact that 
the prefactors in eq.~\eqref{eq:WpsTTTT}, $\Gamma_{T} = 36\sqrt{2}/\pi$
and $A_{T}=64/\pi$ are quite big and of the same order of magnitude. This guarantees large cross-sections and visible effects in SM EFT.
For example, if $C^W=C^{\varphi\Box}>>1$ then the $TTTT$-mode  is again 
bigger than $LLLL$-mode
cross section by a factor of 
$g ^2\Gamma_{T} /\Gamma_{L} = 144\sqrt{2} (G_F m_W^2) \simeq 15$. In this case,
 only 
contributions with \( \mathrm{dim6}^2 \)  in eqs.~\eqref{eq:WpsTTTT}-\eqref{eq:WpsLLLL}  are dominant,  an  approximation  which 
is independent from the cutting-angle $\theta_{cut}$.
%

\begin{figure}
\begin{center}
\includegraphics[scale=0.8]{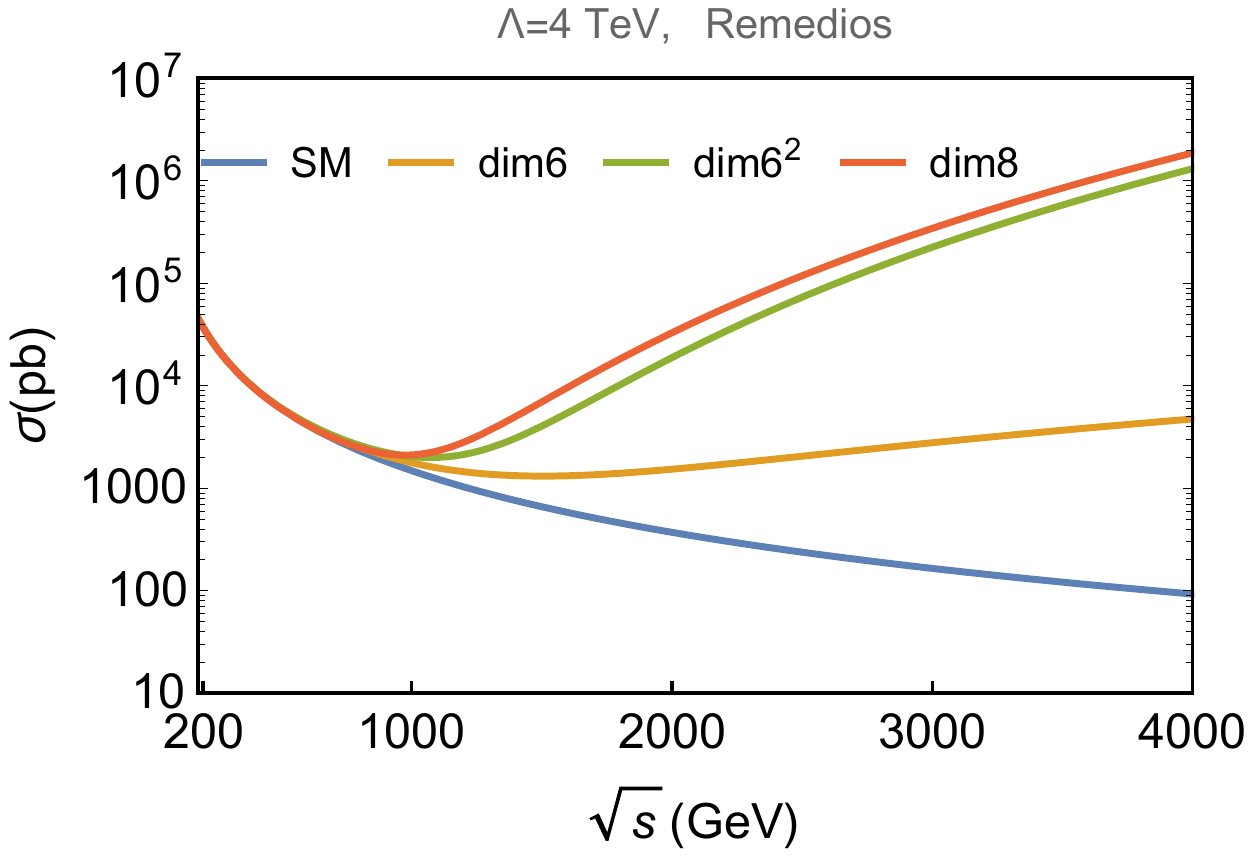} 
\caption{\sl Effects of (dimension-6)$^2$ and dimension-8 operators contributions for $C^W=g_*$, $C^{t0,t1,t2,t10} = g_*^2$, with $g_*=4 \pi$ in the so called Remedios model for $W^+W^+$ elastic scattering. The curves show the 
transversely polarized cross section, $\sigma_{TTTT}$,
 starting from the SM (lower curve), ending at
the full SM+dim6+dim6$^2$+dim8 (upper curve) helicity amplitudes. 
In Remedios the cross-section cannot be 
smaller than the dim6-curve (orange curve).} 
\label{fig:dim8}
\end{center}\end{figure}

We demonstrate these effects in Fig.~\ref{fig:Wp}. We have chosen $\Lambda=4~\text{TeV}$, and,  two non-zero 
Wilson coefficients: $C^W=4\pi$ which affects only the $TTTT$- and $LTLT$-modes and $C^{\varphi\Box} = -4\pi$ which affects only the $LLLL$-mode.\footnote{The minus sign in  $C^{\varphi\Box}$ results in constructive interference.}
The set of input values, although indicative, 
are consistent with the bounds from diboson production
and electroweak fits~\cite{Dawson:2020oco} (see below) when marginalized over other coefficients.
Obviously the appearance of factors $(s/\Lambda^2)^n, n=1,2$  in eqs.~\eqref{eq:WpTTTT}-\eqref{eq:WpLTLT}
turns the  cross sections to  rising. As explained, 
the dominant mode is the $\sigma_{TTTT}$, i.e., the effect of $C^W$ seems the most
promising.   The steep rising of the $LLLL$-cross section for low $\sqrt{s}$ is due to the interference of SM and new physics arising from dimension-6 operators. 
From the right panel of Fig.~\ref{fig:Wp} it is obvious that  
the  $TTTT$ and $LLLL$ cross sections are enhanced by a factor of $\sim$10 much 
before the EFT validity upper bound\footnote{Tree-level unitarity bounds are discussed in detail the in next section.}, 
$s\approx \Lambda^2$, which is encouraging to
make us proceed to a more realistic VBS analysis at the LHC. 
We do this in the next section. 

Furthermore, 
in the case of transverse gauge boson scattering, we considered the effect
 of dimension-8 operators.  It is well known~\cite{Liu:2016idz,Azatov:2016sqh} that dimension-8 operators may dominate (up-to cancellations between coefficients induced by the UV-theory) 
the VBS cross-sections at energies
\begin{equation}
\frac{s}{\Lambda^2} \gg \frac{\bar{g}}{g_*} \;.
\end{equation}
Here we consider the case of strongly coupled (or composite) transverse polarizations of $W$'s, in  the so called \emph{Remedios} models of ref.~\cite{Liu:2016idz}. In these models, $C^W\sim g_*$ and $C^{t0,t1,t2,t10} \sim g_*^2$ 
with the later being the Wilson coefficients of dimension-8 operators defined in  Appendix A.2. There,
we have calculated all dominant helicity amplitudes with leading-$s$ and $s^2$ behaviour. All but one independent helicity amplitudes are masked by 
the $t$-channel photon and $Z$-boson exchange from the SM. The exception is the 
helicity violating amplitude $\mathcal{M}_{++--}$ (or $\mathcal{M}_{--++}$). 
When the coupling 
$g_*$ of  the integrated UV-theory  is strong, i.e. much larger than the 
$SU(2)$-gauge coupling $\bar{g}$, these amplitudes give a large and positive contribution to the 
cross section.  
 We therefore conclude,
 that the inclusion of only the dimension-6 operator $C^W$ is
a \emph{robust conservative limit} to $W^+W^+$ VBS. 
Notice also that (dimension-6)$^2$ contributions in the amplitude 
come with opposite sign to the dimension-8 contributions when 
positivity constraints~\cite{Remmen:2019cyz,Yamashita:2020gtt,deRham:2017zjm} are taken into account, which improves the EFT convergence.\footnote{See~\cite{Ellis:2020ljj} for an interesting example in which the growth with energy, although substantial, does not lead to limitations due to pure EFT convergence. Cases studied are operators that contribute to neutral aTGCs. Such operators start at dimension-8 in SM EFT. Interestingly the identified observables at $e^+e^-$-colliders feature (dimension-8)$^2$ terms suppression, leading to ``genuine'' dimension-8 analysis.}
The situation is clearly explained in Fig.~\ref{fig:dim8}. 

 It is remarkable that in the case of Remedios-like scenarios the estimate on Wilson coefficients, based on power counting as given in ref.~\cite{Liu:2016idz}, does not allow for fulfilment of positivity bounds. This suggests that positivity constraints, eqs.~(6.7)-(6.12) of ref.~\cite{Yamashita:2020gtt}, constitute a significant limitation to models with composite dynamics of transverse electroweak modes. 
 Notice however that these conditions  can still be satisfied if e.g. 
 departures from the power counting estimate is present  in strongly coupled regime such as $C^{t0}=C^{t1}=C^{t10}=|C^W|^2=g_*^2$ and  $C^{t2} \ge 5 g_*^2$.

\section{The realistic study: $pp\to jjW^+W^+$ at the LHC}
\label{sec:anal}

In this section we describe the steps of our analysis  and present numerical results relying on Monte Carlo simulations. The reaction in question is $pp\rightarrow jj W^+W^+\rightarrow jj ll' \nu_l\nu_{l'}$, where $l=e^+,\mu^+$.  
The main goal is to estimate possible effects in same-sign $WW$ scattering in the HL-LHC setup via the EFT approach.  
We analyse in an uncorrelated way one operator at a time while setting all the remaining Wilson coefficients
to zero. Strictly speaking the above choice implicitly assumes certain subset of BSM scenarios where it is valid. Indeed, it constitutes a realistic assumption, e.g. (a) in the case of certain universal models with only bosonic weakly coupled BSM sector \cite{deBlas:2017xtg,Marzocca:2020jze} (given the current experimental constraints on EFT coefficients), (b) scenarios in which  transversal modes of vector bosons  have composite origin~\cite{Liu:2016idz,Marzocca:2020jze}. Nevertheless, the only non-trivial aspect the above simplification misses, is interference terms between different operators. Given the fact that different operators interfere only if they modify the same helicity amplitudes~\cite{Chaudhary:2019aim}, it is significantly limited in our set-up by just looking at eqs~\eqref{amp:ppppTTTT}-\eqref{amp:ppppCPV} of the Appendix~\ref{app:A}. 

 Below 
we focus on largest possible deviations using as our ``benchmarks'' the experimentally allowed  boundary values for each $f_i$ in eq.~\eqref{eq:lagrangian}. We take as the source both the individual-operator-at-a-time limits and those from the global fit for physics scenarios complementarity. In the latter case, correlations between different operators are in principle important. We shall comment on the validity of our one-operator-at-a-time analysis below. Alike, we shall also discuss the effects of background EFT operators.

\subsection{Perturbative Unitarity Bounds and Unitarization issues} 
 \label{subsec:unitarityBounds}
The EFT operators in Table~\ref{tab:ops}  induce $WW\rightarrow WW$ amplitude growth [see eqs~\eqref{amp:ppppTTTT}-\eqref{amp:ppppCPV}] which ultimately leads to violation of probability conservation at some energy scale $\sqrt{s^U}$, a certain $WW$-pair invariant mass, the latter being a function of $f_i$. 
In principle it may happen that $\sqrt{s^U}$ is within the accessible range of $WW$-mass and we found that, in fact, it is the case sometimes. Since predictions of the EFT amplitudes are ill-defined above $\sqrt{s^U}$, the issue has to be addressed in some way in the analysis. 
In this work
we applied additional weights to events above $M_{WW}=\sqrt{s^U}$ in the original non-regularized samples generated with Monte Carlo. The weights are in general operator-dependent. The applied procedure is supposed to ensure that the total $WW$-scattering BSM cross sections
after regularization behave like $1/s$ for $M_{WW} > \sqrt{s^U}$, and so it approximates the principle of constant amplitude, at least after some averaging over the individual helicity combinations~\cite{Kalinowski:2018oxd}. Our choice of unitarization is  often referred to  
in the literature as the (helicity-averaged) ``Kink'' method.\footnote{See Ref.~\cite{Garcia-Garcia:2019oig} for a review 
and comparisons between several unitarization schemes.} We found  the above mentioned weight to be equal
to $(\sqrt{s^U}/M_{WW})^{3.5}$ for operator  $Q_{W}$  and $(\sqrt{s^U}/M_{WW})^{1.5}$ for $Q_{\varphi\Box}$  and $Q_{\varphi W}$.\footnote{ The different exponents follow from the fact that unitarity
is first violated  before the cross section gets dominated by its asymptotic terms.} We refer to such unitarized signal estimate as total BSM signal.

Technically,  $\sqrt{s^U}$ is determined by using the perturbative (tree-level) partial wave unitarity condition~\cite{Jacob:1959at,Barger:1990py}. The statement on $s^U$ from this condition is that for energies $s>s^U$ perturbativity of the EFT necessarily breaks down.
In more detail, the unitarity limit has been determined by studying  all helicity combinations for both  $W^+W^+$ and $W^+W^-$ elastic scattering amplitudes.\footnote{Both channels are governed by the same Wilson-coefficients.} For each helicity amplitude, the first non-vanishing partial wave $\mathcal{T}^{(J)}$ (where always  $J=0,1$ or $2$) is identified and the unitarity bound is found. 
The  scale where unitarity is violated, $\sqrt{s^U}$, is then identified as the lowest value among all such bounds. More explicitly, at tree-level the condition reads (for a detailed discussion  see~\cite{chankowski:notes,Kozow:2019ihv})
\begin{equation}
\sqrt{N_{\lambda_a\lambda_b}N_{\lambda_1\lambda_2}}\left|\mathcal{T}^{(J)}_{\lambda_a\lambda_b;\lambda_1\lambda_2}(s)\right|\leq\frac{1}{2}\;,
\label{eq:unitarityCondition}
\end{equation}
where indices $\lambda_a,\lambda_b$ ($\lambda_1,\lambda_2$) denote outgoing (incoming) helicities, whereas, $N_{xy}=1/2$ for identical particles i.e. $x=y$,  or otherwise  $N_{xy}=1$.  Then the partial wave amplitudes, $\mathcal{T}^{(J)}_{\lambda_a \lambda_b;\lambda_1\lambda_2}(s)$,
 enter in the partial wave expansion as:
\begin{equation}
 \mathcal{M}_{\lambda_a \lambda_b;\lambda_1\lambda_2} \ =\ 16\pi\sum_{J=0}^\infty (2J+1)\mathcal{T}^{(J)}_{\lambda_a\lambda_b;\lambda_1\lambda_2}(s)D^{(J)\ast}_{\lambda_1-\lambda_2, \lambda_a - \lambda_b}(\Omega_{\mathbf{p}_{{(ab)}}}) \;,
\label{eq:partialWaveExpansion}
\end{equation}
with $\Omega_{\mathbf{p}_{{(ab)}}}$ is a solid angle in the direction $\mathbf{p}_{(ab)}=\mathbf{p}_a+ \mathbf{p}_b$, and  $D_{m_J,\lambda}^{(J)}(\Omega_{\mathbf{p}})$ are the  
Wigner-functions satisfying the completeness relation
\begin{equation}
\int{d\Omega_{\mathbf{p}}\,D^{(J')}_{m_J',\lambda}(\Omega_{\mathbf{p}})D^{(J)\ast}_{m_J,\lambda}(\Omega_{\mathbf{p}})=\frac{4\pi}{2J+1}\delta_{J'J}\delta_{m'_Jm_J}}\;.
\label{eq:orthogonalityWigners}
\end{equation}
 We analysed the terms that grow with energy ($\propto f_i$) in which case no Coulomb singularity occurs and correspondingly no phase-space regularization has to be applied (e.g. a cut of $1\deg$ in the forward and backward scattering regions). The results are cross-checked with VBFNLO 1.4.0~\cite{Arnold:2008rz} for operators where direct applicability of the latter tool is possible in the context of  Warsaw basis, obtaining good agreement in the unitarity limits. When applied, the unitarity bound scale, $\sqrt{s^U}$, will be denoted by vertical lines in $M_{WW}$ distributions in figures below. Notice also that the unitarity bounds of VBS processes for dimension-6,8 operators were presented in refs.~\cite{Corbett:2014ora,Almeida:2020ylr}.

We would like to emphasise, that within the EFT approach one does not have knowledge what happens above $s^U$. It could be SM-like perturbative completion or the theory could be non-perturbative. In the former case one expects $\sim1/s$ while in the latter $\sim( \log s)^2$, i.e. the saturation of  Froissart bound~\cite{PhysRev.123.1053}     for the asymptotic behaviour of cross-sections.
	   The Froissart bound has the advantage of working in non-perturbative sense and might be more appropriate in case of $Q_W$ treatment -- the latter is currently poorly constrained. 
	  Applying the Froissart bound to the tail would enhance the signal, although only slightly, as we argue below; 
	  hence our results are being somewhat onto the conservative side.
	  
	  We verified
		that bulk of the SM EFT effect is within the EFT-valid region, i.e. originates from the region $s < s^U$. To this aim, we defined the ``EFT-controlled'' signal estimate~\cite{Kalinowski:2018oxd} and compared it with the total BSM signal (the total BSM signal defined at the beginning of this section). The ``EFT-controlled'' signal estimate is calculated by replacing the generated high-mass tail $M_{WW}>\sqrt{s^U}$ with the one expected in the SM, while taking the EFT prediction for the region $M_{WW}<\sqrt{s^U}$. Hence, the ``EFT-controlled'' defines a signal originating uniquely from the operator within its (maximal) range of EFT validity. In turn, comparison between the total BSM signal estimate with the ``EFT-controlled'' signal allows for a verification of the significance of the tail region -- the conclusions based on EFT are reliable only if bulk of the BSM signal is in the ``EFT controlled'' region~\cite{Kalinowski:2018oxd}. 
	  In Sec.~\ref{subsec:mainops} we show that our results are approximately independent of how the contribution in the region above the unitarity bound, $s>s^U$, is estimated, i.e. this issue should be a secondary effect; quantitatively, we have checked that the total BSM signal and the ``EFT-controlled'' estimate are statistically consistent within 2$\sigma$ for all the operators and Wilson coefficients studied in sec.\ref{subsec:mainops}, with the exception of $f_W=1$ TeV$^{-2}$ (discussed further therein).

Although it is well known that non-unitarized results do not directly have a physics interpretation, they could be safely taken as overestimated upper bounds on any unitarized results.  For this reason, agreement between non-unitarized results and the ``EFT-contolled'' estimate implies little dependence on unitarization details. We did the corresponding comparison in addition and we observe consistency within $2\sigma$ between non-unitarized results and the ``EFT-controlled'' results for all the cases (except $f_W=$1 TeV$^{-2}$). 
Therefore we choose to use the total BSM signal defined at the beginning of this section (popularly referred to as ``Kink'' method) to obtain our central results throughout this work.

	  Note also, that positivity bounds are not harmed by considering constant amplitudes for the asymptotic behaviour of cross-sections in our analysis.

\subsection{Method and Kinematic variables}

For the following analysis two samples of $6\times 10^5$ events consistent with the VBS topology for the process $pp\rightarrow jj\mu^+\mu^+\nu\nu$ have been generated, each corresponding to a preselected arbitrary value of $f_W$ or $f_{\varphi\Box}$ coefficient, in MadGraph5\_aMC@NLO~\cite{Alwall:2014hca} v2.6.2 at LO at 14 TeV $pp$ collision energy. Different $f_i$-values were obtained by applying weights to generated events, using the
reweight command in MadGraph. The value $f_i=0, \forall i$, represents the SM predictions for
each study. Results for the remaining relevant operators, i.e. $Q_{\varphi D}, Q_{\varphi W}$, were obtained using the re-weight command with  $Q_{\varphi\Box}$ sample.\footnote{ Samples for $Q_{W}$ were generated separately due to large reported uncertainties when reweighting from $Q_{\varphi\Box}$.}

The SmeftFR code~\cite{Dedes:2019uzs} v2.01 (based on FeynRules~\cite{Alloul:2013bka})  was used to generate the UFO file~\cite{Degrande:2011ua} with an 
input parameter scheme $\{G_F,m_W,m_Z,m_h\}$,  in SM EFT. Cross sections at the output of MadGraph are multiplied by a factor 4x to account for all the lepton combinations in the final state. Hadronization is done with Pythia v8.2~\cite{Sjostrand:2006za,Sjostrand:2007gs}, run within MadGraph. Reconstruction level is generated via the MadAnalysis5~\cite{Conte:2012fm} v1.6.33 package (available within MadGraph). The FastJet~\cite{Cacciari:2011ma} v3.3.0 package is used with the jet clustering \verb+anti-kT+ algorithm with \verb+radius=0.35+ and \verb+ptmin=20+. Finally, the detector efficiencies are set to 100\%. 

The SM process $pp \to jj\ell^+\ell^+\nu\nu$ is treated as the irreducible background, while the ``signal'' is defined as the enhancement of the event yield relative to the SM prediction in the presence of a given operator $Q_i$. None of the reducible backgrounds are simulated. The reason is that reducible backgrounds, as we learn from e.g. Fig.~3 of a recent study~\cite{Sirunyan:2020gyx}, roughly doubles the total statistics overall in the VBS fiducial region and is mostly concentrated at low mass (both dilepton and dijet).  Since VBS-related operators modify mainly the opposite end of the spectrum, reducible backgrounds probably will not be crucial to have an estimate of the possible effects. Yet another aspect is that additional operators might also modify the reducible backgrounds in some unforeseen ways. But such potential effects will need to be determined experimentally from other studies, in which those processes are not backgrounds but signals. We do not address this issue in our paper.

Following ref.~\cite{Kalinowski:2018oxd}, the event selection criteria consist of requiring at least two reconstructed jets and exactly two leptons (muons or electrons) satisfying the following conditions: $M_{jj} > 500$~GeV,  $\Delta\eta_{jj} > 2.5$, $p_T^{~j} > 30 $~GeV, $|\eta_j| <5$, $p_T^{\ell} >25$~GeV and $|\eta_\ell| < 2.5$, being $\eta_{j,\,\ell}$ the pseudorapidity of jets $j$ or leptons $\ell$, respectively. 
The total BSM signal significances are computed as the square root of a $\chi^2$ resulting from a bin-by-bin comparison of the event yields in the binned distributions of different kinematic observables. Moreover, event distributions are always normalized to HL-LHC luminosity, i.e. 3000 fb$^{-1}$.

For each benchmark value of $f_{i=W,\varphi\Box,\varphi D,\varphi W}$, the signal significance is assessed by studying the
distributions of a large number of kinematic variables. These are: 
\begin{equation} 
\def\arraystretch{1.5}
\begin{array}{l}
m^{jjll}, m^{ll}, m^{jj}, p_T^{j1}, p_T^{j2}, p_T^{l1}, p_T^{l2}, \\
\eta^{j1}, \eta^{j2}, \eta^{l1}, \eta^{l2}, d\,\eta^{j}, d\,\phi^{j}, d\,\phi^{l}, \\
R_{p_T} \equiv p_T^{~l1}p_T^{~l2}/(p_T^{~j1}p_T^{~j2}),  \\
M_{o1} \equiv \sqrt{(|\vec{p}_T^{~l1}|+|\vec{p}_T^{~l2}|
+|\vec{p}_T^{~miss}|)^2 - (\vec{p}_T^{~l1}+\vec{p}_T^{~l2}+\vec{p}_T^{~miss})^2}, \\ 
M_{1T}^2\equiv \left(\sqrt{\left(m^{ll}\right)^2 +\left(\vec{p}^{~l1}\right)^2+\left(\vec{p}^{~l2}\right)^2}+\left|\vec{p}_T^{~miss}\right|\right)^2 - \left(\vec{p}^{~l1}+\vec{p}^{~l2}+\vec{p}_T^{~miss}\right)^2,
\end{array}
\label{eq:listOfVariables}	
\end{equation}
where $l1(2)\equiv$(sub) leading lepton; $j1(2)\equiv$(sub) leading jet; $m$ invariant mass; $p_T^{(miss)}\equiv$ (missing) transverse momentum; $\eta\equiv$ pseudo-rapidity, $\phi\equiv$ azimuthal angle. 
 Some of these variables are well known to be VBS-blind, but in the context of ``background operator'' they may still be useful.
We found both angular variables, $\eta$s and $\phi$s, as well as $p_T$s involving jet(s), to be at most sub-leading in sensitivity. Moreover, all $p_T$s, $m$s and $M$s were found to be at most sub-leading in sensitivity in case of $Q_{\varphi\Box}$ and $Q_{\varphi D}$. 
The most sensitive variables were found to be: $p_T^{l1}$ for $Q_{W}$; $M_{o1}$ for $Q_{\varphi W}$; and $R_{p_T}$ for $Q_{\varphi\Box},Q_{\varphi D}$.

We considered one-dimensional distributions of single variable. Each distribution is divided into 10 bins, arranged so that the Standard
Model prediction in each bin is never lower than 2 events. Overflows were always included in the respective highest bins. Presented results and conclusions are 
always based upon the most sensitive variables, which is, in general, an operator dependent outcome.

The total BSM signal significance expressed in standard deviations ($\sigma$)
is defined as the square root
of a $\chi^2$ resulting from comparing the bin-by-bin event yields:

\begin{equation}
\chi^2 = \sum_i (N^{BSM}_i - N^{SM}_i)^2 / N^{SM}_i.
\end{equation}

\begin{table}[t]
 \vspace{0.5cm}
 \centering
 \begin{tabular}{||cc|c||cc|c||}

		\hline
		\hline
   \multicolumn{2}{||c|}{$\psi^2\varphi^3$~\cite{Ellis:2018gqa}}& $\sigma$ & \multicolumn{2}{c|}{$\psi^2X\varphi$~\cite{Ellis:2018gqa}}& $\sigma$ \\
	\hline
  $f_{u\varphi}$ & $[-120.,-36.]\times y_u$ & 0.027 & $f_{uG}$ & $[+5,+18.]\times y_u$ & $5.5\times 10^{-3}$ \\
	$f_{d\varphi}$ & $[+3.,+7.9]\times y_d$   & 0. &          &                       &      \\
	\hline\hline
	
 \multicolumn{2}{||c|}{$\psi^2\varphi^2D$~\cite{Dawson:2020oco}}& $\sigma$ & \multicolumn{2}{c|}{$(\bar{L}L)(\bar{L}L)$~\cite{Domenech:2012ai,Sirunyan:2017ygf}}& $\sigma$ \\ \hline
	
	$f_{\varphi q}^{(1)}$ & $[-0.23,+0.12]$   & 0.46 & $f_{qq}^{(1)}$ &$[-0.028,+0.057]$ & 1.1 \\ 
	$f_{\varphi q}^{(3)}$&$[-0.18,+0.17]$ & 5.7     &          &                       &      \\
	$f_{\varphi u}$      &$[-0.79,+0.54]$&0. &          &                       &      \\
	$f_{\varphi d}$       &$[-0.81,+0.13]$ &0. &          &                       &      \\
	
		\hline\hline
	\multicolumn{2}{||c|}{$(\bar{R}R)(\bar{R}R)$~\cite{Domenech:2012ai}}& $\sigma$ & \multicolumn{2}{c|}{$(\bar{L}L)(\bar{R}R)$~\cite{Domenech:2012ai}}& $\sigma$ \\ \hline
	
	$f_{uu}$ &$[-0.1,+0.23]$ & 0.&$f_{qu}^{(1)}$ &$[-0.35,+0.35]$& 0. \\ 
	$f_{dd}$ &$[-0.31,+0.44]$ & 0.& $f_{qu}^{(8)}$ &$[-0.5,+1.]$  		& 0. \\			
	$f_{ud}^{(1)}$ &$[-0.44,+0.44]$& 0. & $f_{qd}^{(1)}$ &$[-0.59,+0.59]$& 0. \\
	$f_{ud}^{(8)}$ &$[-0.59,+1.56]$& 0. & $f_{qd}^{(8)}$ & $[-1.,+1.56]$& 0. \\
	 \hline
	\hline
						%
						%
 \end{tabular}
 \caption{\it Compilation of all the experimental limits coming from global fits on dimension-6 background operators used in this work, in TeV$^{-2}$ units, and maximal effect in standard deviations ($\sigma$) in the $W^+W^+$ scattering process from each operator separately. Flavor assumptions are implicit and follow the references quoted. In particular, ``$\times y$" corresponds to the Minimal Flavor Violation assumption.}
 \label{tab:bgLimitsCompilation}
\end{table}

\subsection{Background Operators analysis}

Before moving to the analysis of operators in Table~\ref{tab:ops} that modify the $WW\rightarrow WW$ reaction (VBS operators), we comment on the role of the dimension-6 interactions that, although do not affect the sub-process $WW\rightarrow WW$ directly, may contribute at parton level to $pp\rightarrow jj ll' \nu_l \nu_l'$ (background operators). 
This is particularly relevant for consistency of the discussion of potential effects of VBS operators when $f_i$ is constrained by the global fit, but also for a more complete description in general. 
 We examined possible contributions (after the VBS cuts are applied) operator-by-operator, based on the bounds reported in~\cite{Dawson:2020oco,Ellis:2018gqa} (non-4-fermion operators, the global more permissive constraints are implicitly understood below) and~\cite{Domenech:2012ai,Sirunyan:2017ygf} (4-fermion operators) (with consistently applied flavor assumptions). The so-called ``dipole'' operators, i.e. $Q_{uW},Q_{uB},Q_{dG}, Q_{dW},Q_{dB}$, are not present in the quoted references. They are however commonly claimed to be strongly constrained~\cite{Ellis:2018gqa,falkowski:notes}. We checked that among these, only $Q_{uW},Q_{dW},Q_{dG}$ would lead to noticeable effects for a representative $f_i= 1$ TeV$^{-2}$. Importantly however, they also lead to commonly distinctive kinematic features compared to VBS operators, and hence 
 they are in principle easy to be separated 
 as part of ``background'' effects (discussed in detail below). We have limited our examination to non-leptonic operators, because  leptonic operators would contribute only at loop-level to the process $pp\rightarrow jj W W$, and it is known that the on-shell projection of the outgoing $W$'s can be defined in a gauge invariant way and constitute a fine ($\lesssim5\%$ error) approximation~\cite{Ballestrero:2017bxn,Ballestrero:2020qgv}. Moreover, in this work we do not consider CP- or B-violating operators. Out of the remaining, coefficients accompanied the operator $Q_{G}$ are known to be constrained at  $O(0.01)$ TeV$^{-2}$ via multi-jet channels~\cite{Krauss:2016ely}, whereas, $Q_{\varphi}$ does not contribute to the tree-level amplitude. We also omit $Q_{\varphi G}$ as it is constrained by gluon fusion to Higgs-boson at loop-level~\cite{Falkowski:2019hvp}. In addition to this, we found no sensitivity to the 4-fermion operators in the ``$(\bar{L}R)(\bar{R}L)$'' and ``$(\bar{L}R)(\bar{L}R)$'' classes even for generic BSM coupling in the strong interaction regime.
The experimental bounds on the remaining background operators are compiled in Table~\ref{tab:bgLimitsCompilation}. One can see that strong suppression is hidden in the Yukawa factors for the $Q_{u\varphi},Q_{d\varphi},Q_{uG}$ interactions (Minimal Flavor Violation hypothesis is assumed throughout  the analysis). Since the light quarks play (by far) the leading role in our reaction, the Wilson coefficients are effectively of, at least, 1-loop order in this case.

As far as non-4-fermion operators are concerned, we found that the current limits on all background dimension-6 operators yield cross-sections consistent with the SM prediction, with the exception of 
\begin{equation}
Q_{\varphi q}^{(3)} \ = \ (\varphi^\dagger i \overleftrightarrow{D}_\mu^I \varphi)\:(\bar{q} \tau^I \gamma^\mu q)\;,
\label{eq:Qphiq3}
\end{equation}
that yielded systematic discrepancy of around $\sim5\%$. Detailed analysis revealed that this operator gives a up to $\approx 5.5\sigma$ potential discrepancy with SM at the HL-LHC. The distributions in $M_{WW}$ and $p_{T}^{l1}$ are shown in Fig.~\ref{fig:distrosGlobalCphiq3} for $f_{\varphi q}^{(3)}=+0.17$ (the negative boundary value $f_{\varphi q}^{(3)}=-0.18$ is very similar). 

\begin{figure}[ht]
\centering \includegraphics[scale=0.5]{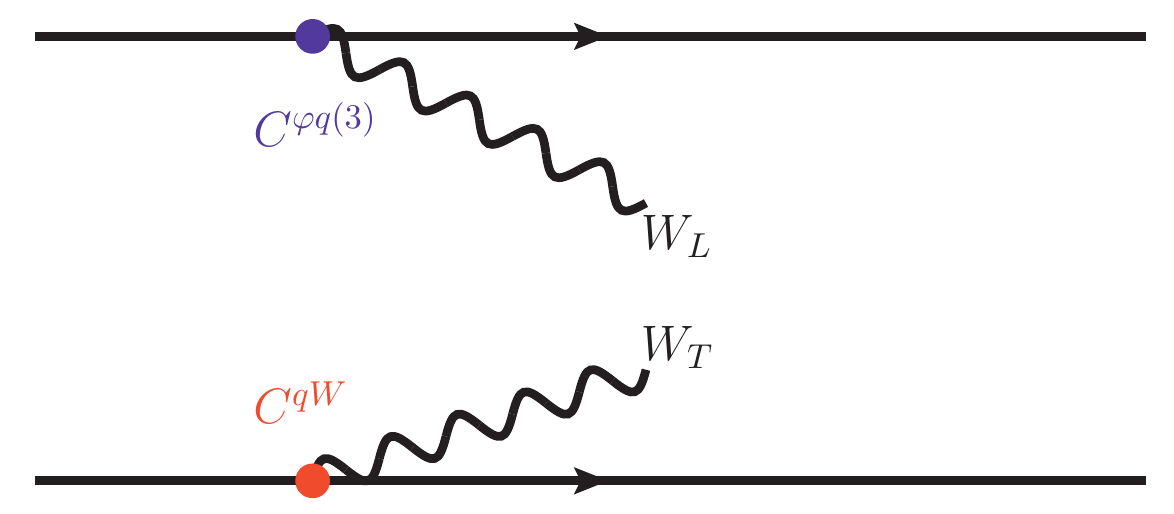} 
\caption{\sl Examples of  background operators indicated by the associated Wilson-coefficients.}
\label{fig:phiq3}
\end{figure}

\begin{figure}[t]
\begin{center}
\mbox{
\includegraphics[scale=0.6]{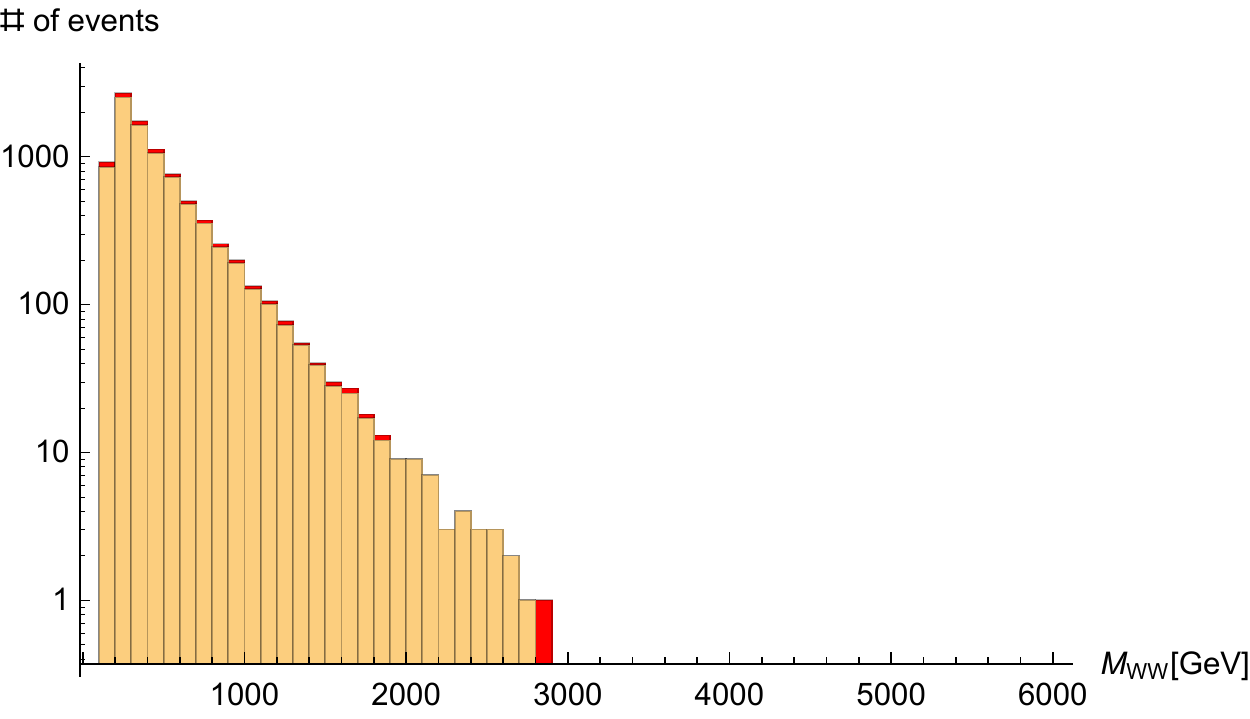} 
\includegraphics[scale=0.6]{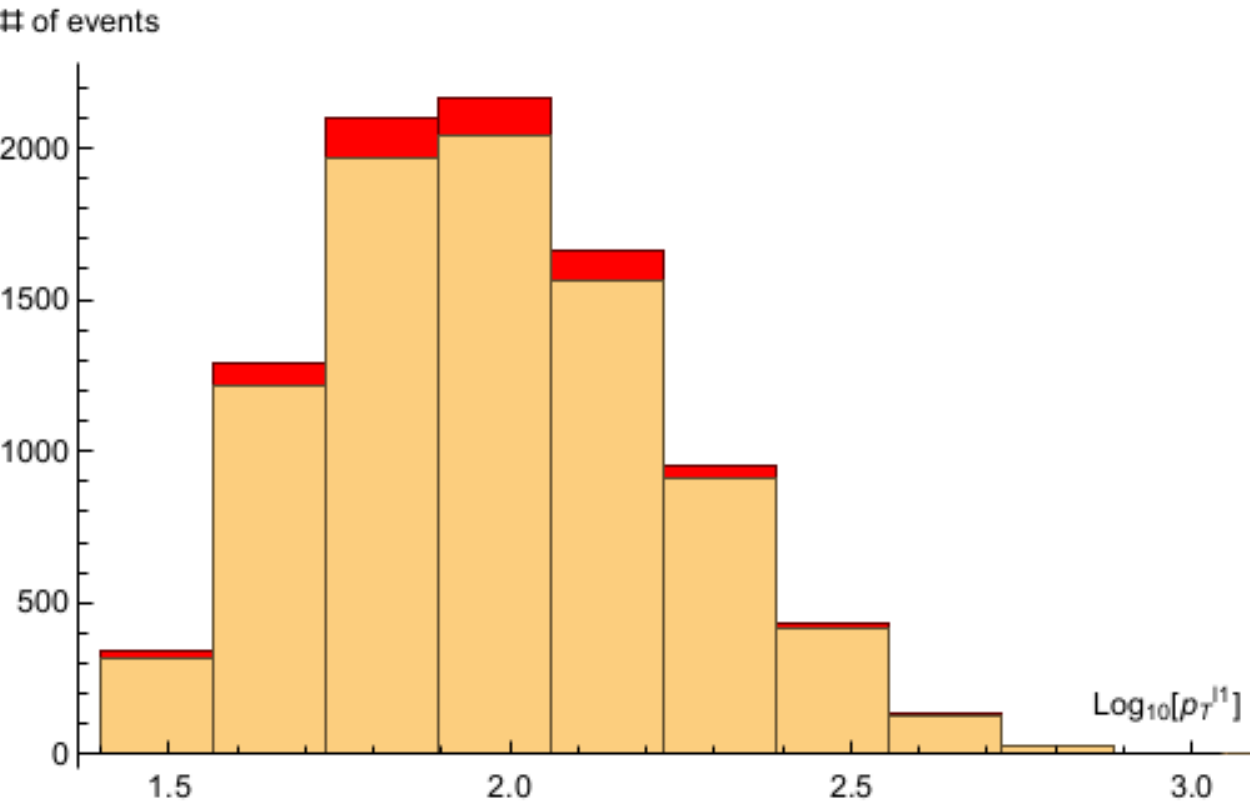} 
}
\end{center}
\caption{\sl The distributions in $WW$ system invariant mass $M_{WW}$ (left) and the most sensitive observable (here $p_{T}^{l1}$) for $f_{\varphi q}^{(3)}=+0.17$ compared to the SM case. The BSM signal estimate is in red and the SM in yellow.}
\label{fig:distrosGlobalCphiq3}
\end{figure}

One can observe that this background operator [see Fig.~\ref{fig:phiq3}] has distinct dynamics compared to the VBS operators (presented subsequently). In particular, the contributions to $\chi^2$ are uniformly distributed among all the bins ($p_T^{l1}$) and a similar feature holds for the $M_{WW}$ distribution. That is expected, because this background operator has nothing to do with  $WW$-mass,  but as shown in  Fig.~\ref{fig:phiq3} it mainly contributes to the hard scattering for
longitudinal $W$-bosons, so the energy dependence measured in  $WW$-processes should be different.

It shows that even when the background operator effect is large, it can be easily identified and disentangled by studying kinematic distributions -- if one gets more than 5 $\sigma$ between $Q_{\varphi q}^{(3)}$  and the SM, then one is likely also to get a 5 $\sigma$ difference between the predicted shapes for $Q_{\varphi q}^{(3)}$ and any VBS operator. More quantitatively,  we found e.g. that $p_T^{j1}$ is the most sensitive variable in the case of $Q_{\varphi q}^{(3)}$\footnote{Notice, it involves a contact interaction between two quarks
and two vectors, so it contributes to a hard scattering of qW.}, while it is by far sub-leading in sensitivity for the VBS operators.
 Moreover, the weak sensitivity of jet-$p_T$s  also holds for the leading dimension-8 operators~\cite{Kozow:2019txg} 
 usually considered  as the leading effects in phenomenological/experimental analyses for VBS processes so far. On the other hand, the highest sensitivity to $p_T^{j1}$ also holds for the relevant dipole operators, as well as for the operator $Q_{\varphi ud}$, which is not studied in the ``bounds" literature either; this constitutes the distinctive dynamics mentioned above for the dipole operators.

Concerning the potential effects in 4-fermion dimension-6 operators, we checked that the limits reported in~\cite{Domenech:2012ai} were already sufficient to claim negligible effects these operators might at most generate. The exception was the left currents operator $Q_{qq}^{(1)}$ for which the search in the more up-to-date ref.~\cite{Sirunyan:2017ygf} was used to derive the factor by which the limits reported in~\cite{Domenech:2012ai} improve. We found the factor to be conservatively estimated by $\sim 3$x, and using the improved constraints concluded that the operator might generate only negligible effects. Note that the other non-leptonic operator in the $(LL)(LL)$ category, $Q_{qq}^{(3)}$, is identical to $Q_{qq}^{(1)}$ assuming flavor-diagonal Wilson coefficients.

\subsection{Main operators analysis}
\label{subsec:mainops}
 From the previous subsection, we conclude that in practice it suffices to parametrize potential NP effects  from Wilson coefficients associated with  the main dimension-6 operators listed in Table~\ref{tab:ops}.  i.e. the operators that modify the $WW\rightarrow WW$ sub-process. By saying this we exclude from our discussion 
CP-Violating operators, $Q_{\widetilde{W}}$ and $Q_{\varphi\widetilde{W}}$~\footnote{The operator $Q_{\varphi \widetilde{W} B}$ does not enter to the leading amplitudes, see formulae in Appendix~\ref{app:1}.} for two reasons: first, because their contribution in the cross section $WW\to WW$  is no different 
than the CP-Conserving $Q_W$ and $Q_{\varphi W}$, and second,
 these operators are (usually) much more constrained than their 
 CP-Conserving cousins.
Therefore, we focus on the CPC operators arranged in Table~\ref{tab:ops}
for the numerical simulations to follow.

\begin{figure}[h]
\begin{center}
\mbox{
\includegraphics[scale=0.6]{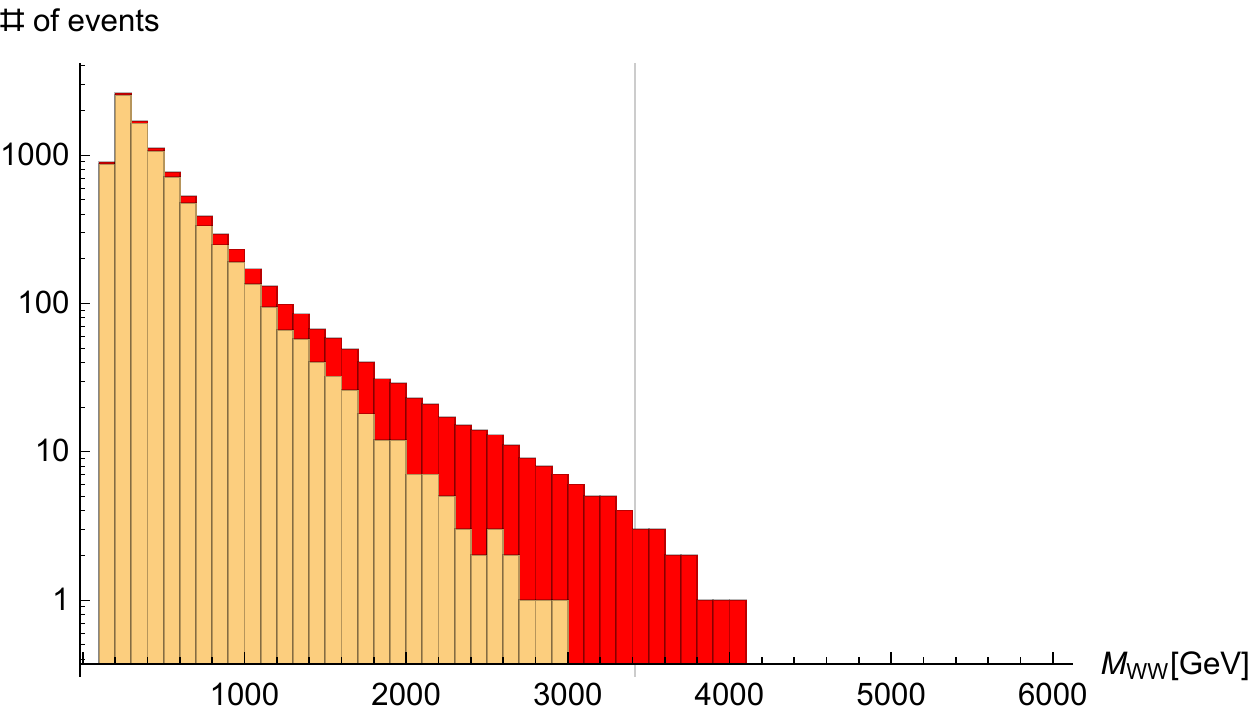} 
\includegraphics[scale=0.6]{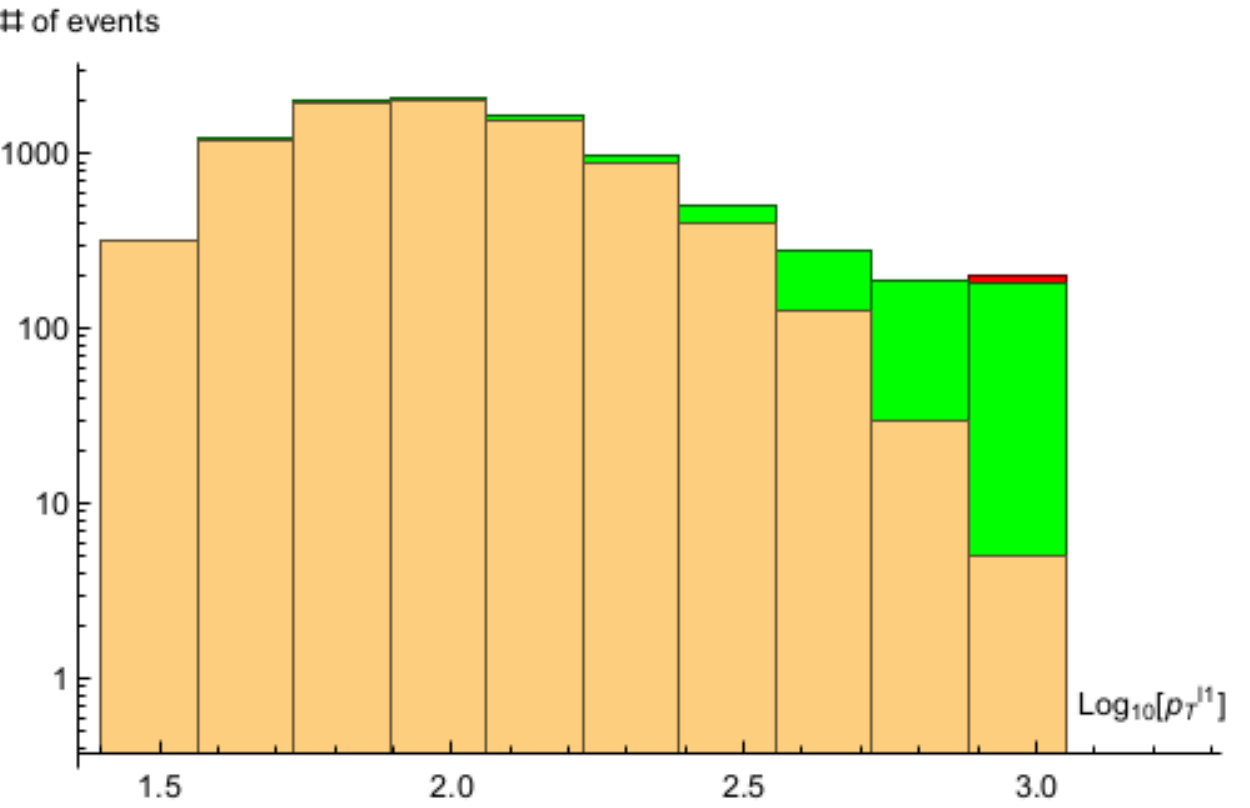} 
}
\end{center}
\caption{\sl Distributions in $WW$ pair invariant mass $M_{WW}$ (left) and the most sensitive observable (right, here $p_{T}^{l1}$) , for $f_{W}=+0.36$ TeV$^{-2}$ compared to the SM case. The total BSM signal estimate is in red and the SM in yellow. Notice, that in the right plot shown is the ``EFT-controlled'' signal estimate in green (almost identical to the red one); normalized to HL-LHC. The vertical line (left) denotes the scale above which the partial wave unitarity condition is violated (see Sec.~\ref{subsec:unitarityBounds}).}
\label{fig:distrosIndividualCWgeq}
\end{figure}

\begin{figure}
\begin{center}
\mbox{
\includegraphics[scale=0.6]{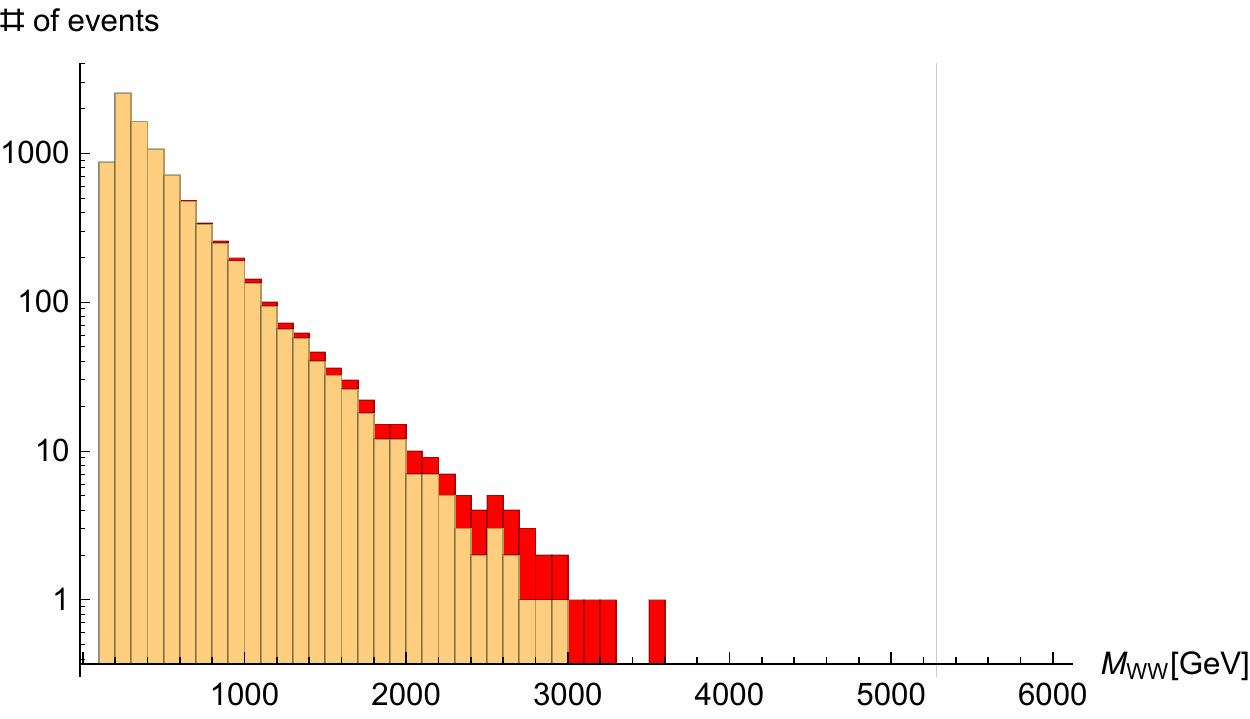} 
\includegraphics[scale=0.6]{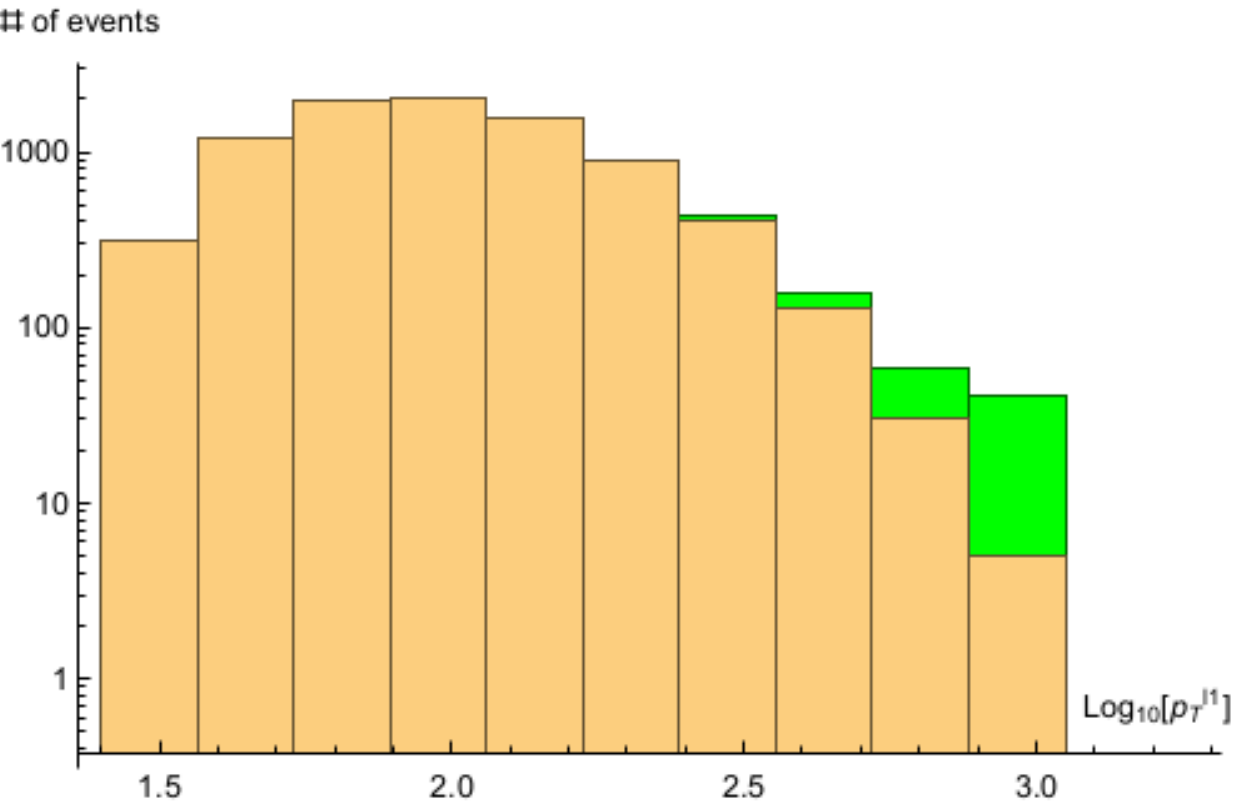} 
}
\end{center}
\caption{\sl Like Fig.~\ref{fig:distrosIndividualCWgeq} for $f_{W}=-0.15$ TeV$^{-2}$.}
\label{fig:distrosIndividualCWleq}
\end{figure}

We begin by examining the possible effects at the HL-LHC assuming the bounds on EFT coefficients come from the individual-operator-at-a-time analysis, based on the recent~\cite{Dawson:2020oco}; we quote the relevant constraints indicated as ``individual" in Table~\ref{tab:limits}. 
\begin{table}[h]%
\begin{center}
\begin{tabular}{|c||c|c|c|c|} \hline
     & $f_{W}$ & $f_{\varphi\Box}$ & $f_{\varphi D}$ & $f_{\varphi W}$ \\ \hline ``individual'' & [-0.15,+0.36] & [-0.44,+0.52]  & [-0.025,+0.0015] & [-0.014,+0.0068] \\ \hline
``global'' & [-1.3,+1.1] & [-3.4,+2.4] &[-2.7,+1.2] & [-0.14,+1.6]\\ \hline
\end{tabular}
\end{center}
\caption{\sl Experimental constraints on the subset of operators modifying the process $W^+W^+\rightarrow W^+W^+$: based on the individual-operator-at-a-time or global marginalized fit analyses; from~\cite{Dawson:2020oco}.}
\label{tab:limits}
\end{table}
The only $f_i$ bounded by as large as $O(0.1)$ TeV$^{-2}$ are $f_{\varphi\Box}$ and $f_{W}$, the remaining two $f_{\varphi D}$ and $f_{\varphi W}$ are bounded by  $O(0.01)$ TeV$^{-2}$ (or stronger, depending on the sign) and give null discrepancies with the SM. In fact, at the boundaries $f_{\varphi\Box}\approx\pm0.5$ yields negligible effects as well. On the other hand, both boundaries on $f_W$ allow for large effects. The distributions $M_{WW}$ (left) and the total BSM signal in the most significant variable $p_T^{l1}$ (right), are shown in Fig.~\ref{fig:distrosIndividualCWgeq} for $f_W=+0.36$ TeV$^{-2}$ and in Fig.~\ref{fig:distrosIndividualCWleq} for $f_W=-0.15$ TeV$^{-2}$.
 Notice, that in the right plots the total BSM signal is drawn with red and the ``EFT-controlled'' signal with green, and that the red histogram is almost entirely covered by the green one, as anticipated in Sec.~\ref{subsec:unitarityBounds}.

Next, we consider bounds on $f_i$ from the global fit (indicated with ``global" in Table~\ref{tab:limits}), again based on recent literature~\cite{Dawson:2020oco}. 
One can observe that  constraints here are considerably more relaxed than the ``individual" ones  allowing for $f_i$ as large as $O(1.)$~TeV$^{-2}$. It is due to correlations between various Wilson coefficients that contribute to the same observables used to determine the bounds. Therefore such bounds could be understood as complementary and in principle saturated in UV models that normally allow for a plethora of dimension-6 coefficients without too large hierarchies in their magnitudes; e.g. one can construct such models exploiting the UV vs tree-level matching dictionary~\cite{deBlas:2017xtg} with a large number of different heavy species.
As no correlation matrix is provided in ref.~\cite{Dawson:2020oco}, in what follows we examine the four VBS-operators in a simplified way, i.e. one-by-one,  and describing qualitatively the effect of correlations below.
 
We take an estimate on correlations from ref.~\cite{Ellis:2018gqa}, which corresponds to similar physics assumptions. One can see that the correlations  among the four relevant operators are mostly negligible. They are however somewhat mild between the following pairs: 
$(Q_{\varphi\Box},Q_{\varphi W})$, $(Q_{\varphi\Box},Q_{W})$, $(Q_{\varphi D},Q_{W})$. Since,
as we show in the Appendix, the operators in each pair modify distinct helicities of the $WW\rightarrow WW$ sub-process, the corresponding effects do not interfere.
As a result, to account for correlations corresponds to adding together the effects presented in this work of each operator,  with appropriate weights from the correlation matrix. The most sensitive variables in~\eqref{eq:listOfVariables} for each operator feature increase of number of events in the higher  (relevant) bins independently of the sign of $f_i$.  Moreover, notice that the account for mild correlation with $Q_{W}$ may affect the estimated effects because this operator is by far dominant in our analysis, and in this sense the results presented below for $Q_{\varphi\Box}$ and $Q_{\varphi D}$ should be regarded as conservative ones.
%

\begin{figure}[h]
\begin{center}
\mbox{
\includegraphics[scale=0.6]{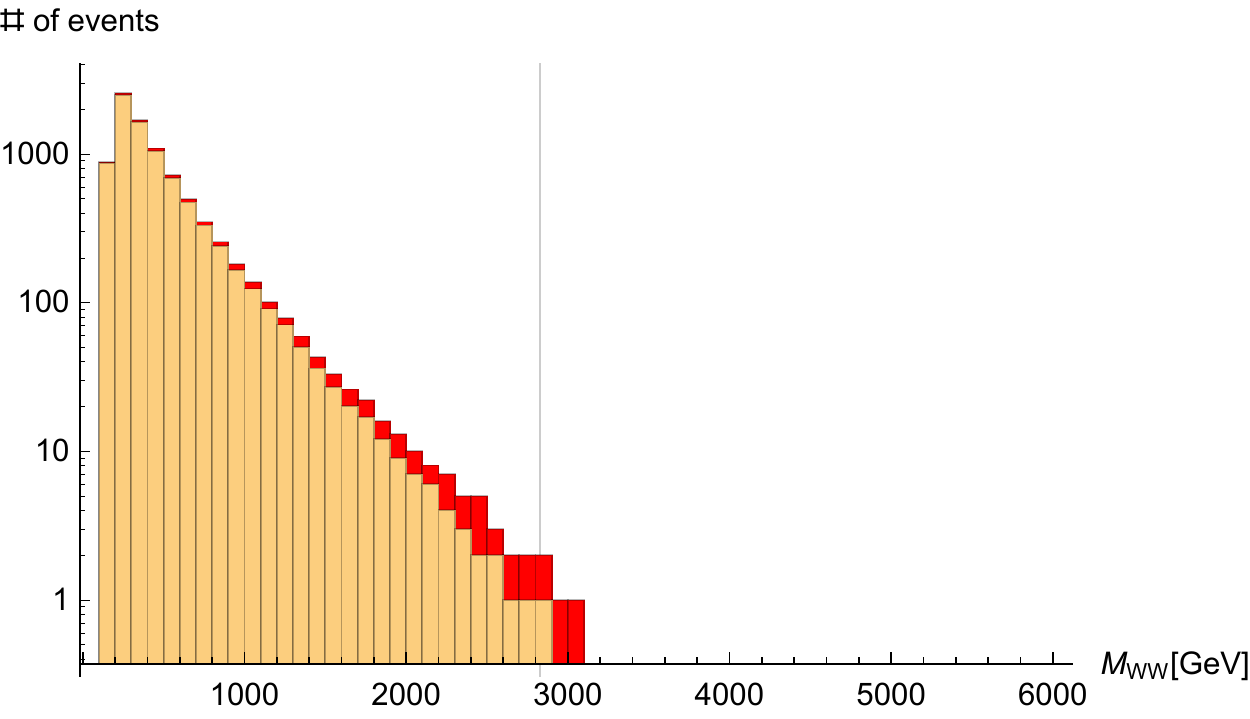} 
\includegraphics[scale=0.6]{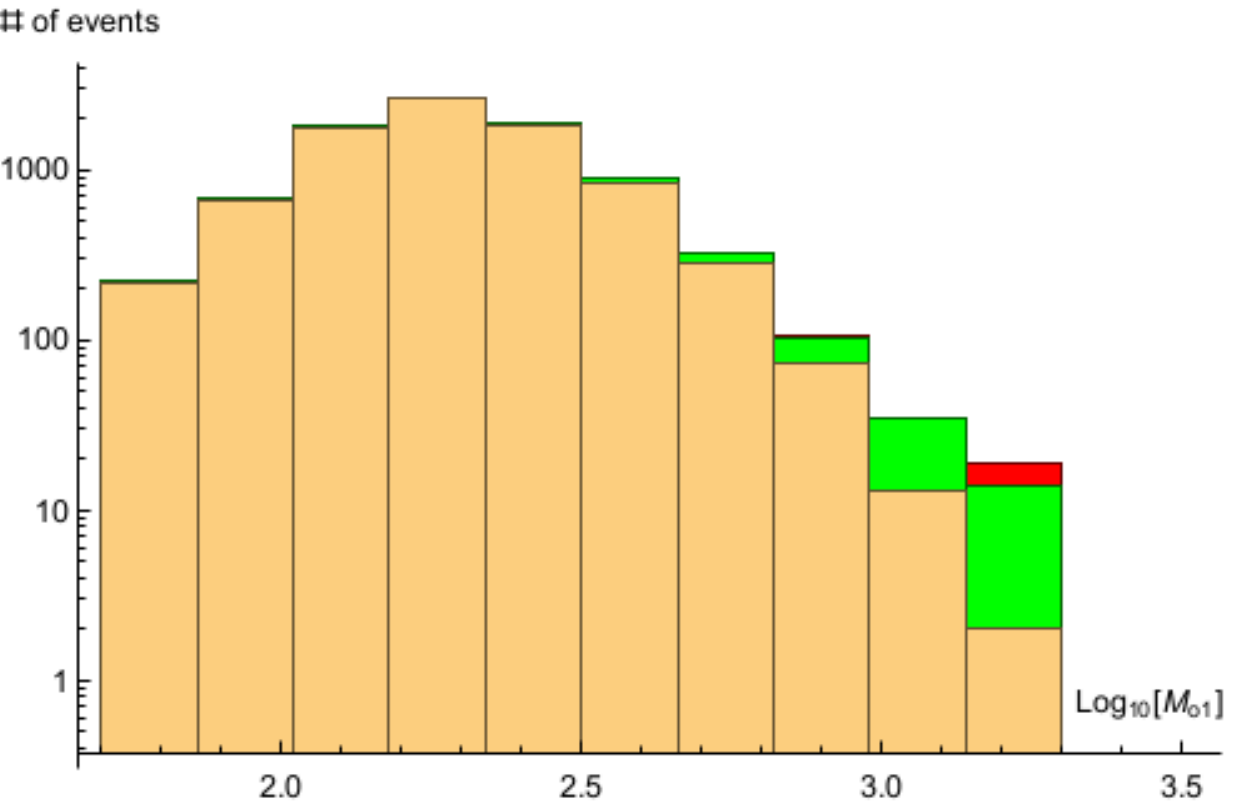} 
}
\end{center}
\caption{\sl Like in Fig.~\ref{fig:distrosIndividualCWgeq} for  $f_{\varphi W}=+1.6$ TeV$^{-2}$.}
\label{fig:plotHisto19SMcphiW_1_6}
\end{figure}
\begin{figure}[h]
\begin{center}
\mbox{
\includegraphics[scale=0.6]{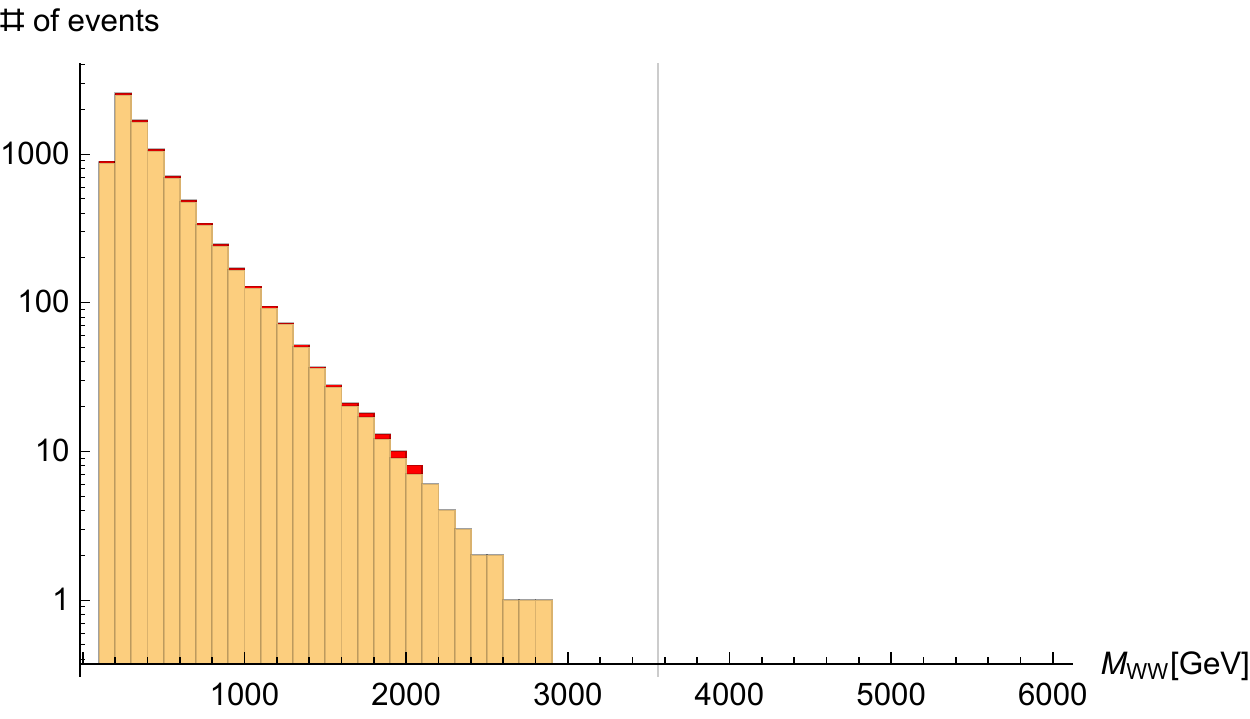} 
\includegraphics[scale=0.6]{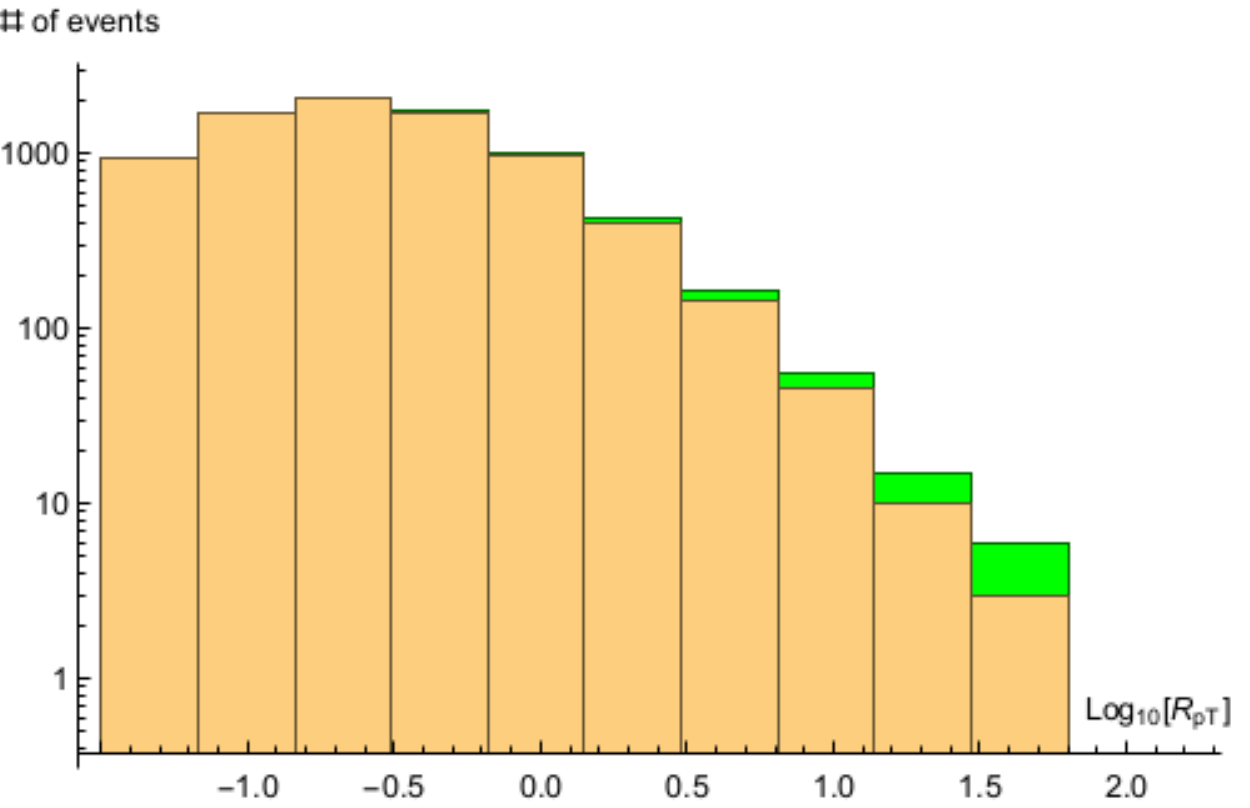} 
}
\end{center}
\caption{\sl Like in Fig.~\ref{fig:distrosIndividualCWgeq} for  $f_{\varphi D}=-2.7$ TeV$^{-2}$.}
\label{fig:plotHisto19SMcphiD_m2_7}
\end{figure}
\begin{figure}[h]
\begin{center}
\mbox{
\includegraphics[scale=0.6]{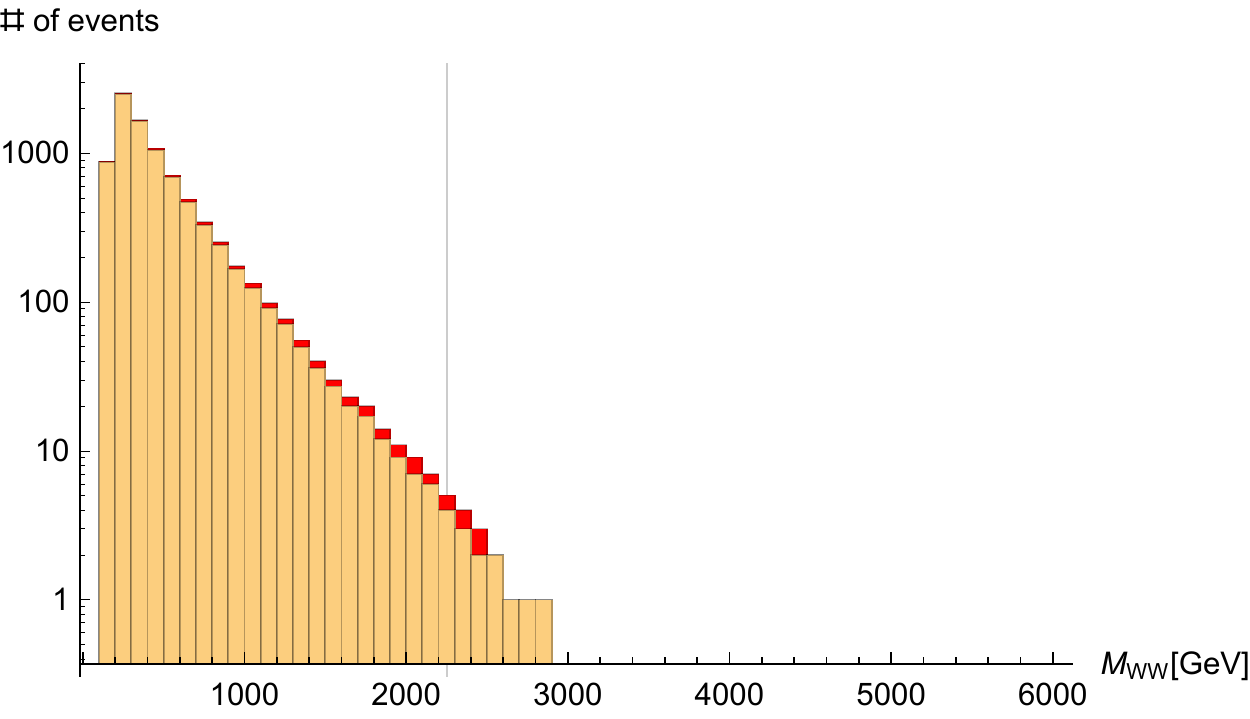} 
\includegraphics[scale=0.6]{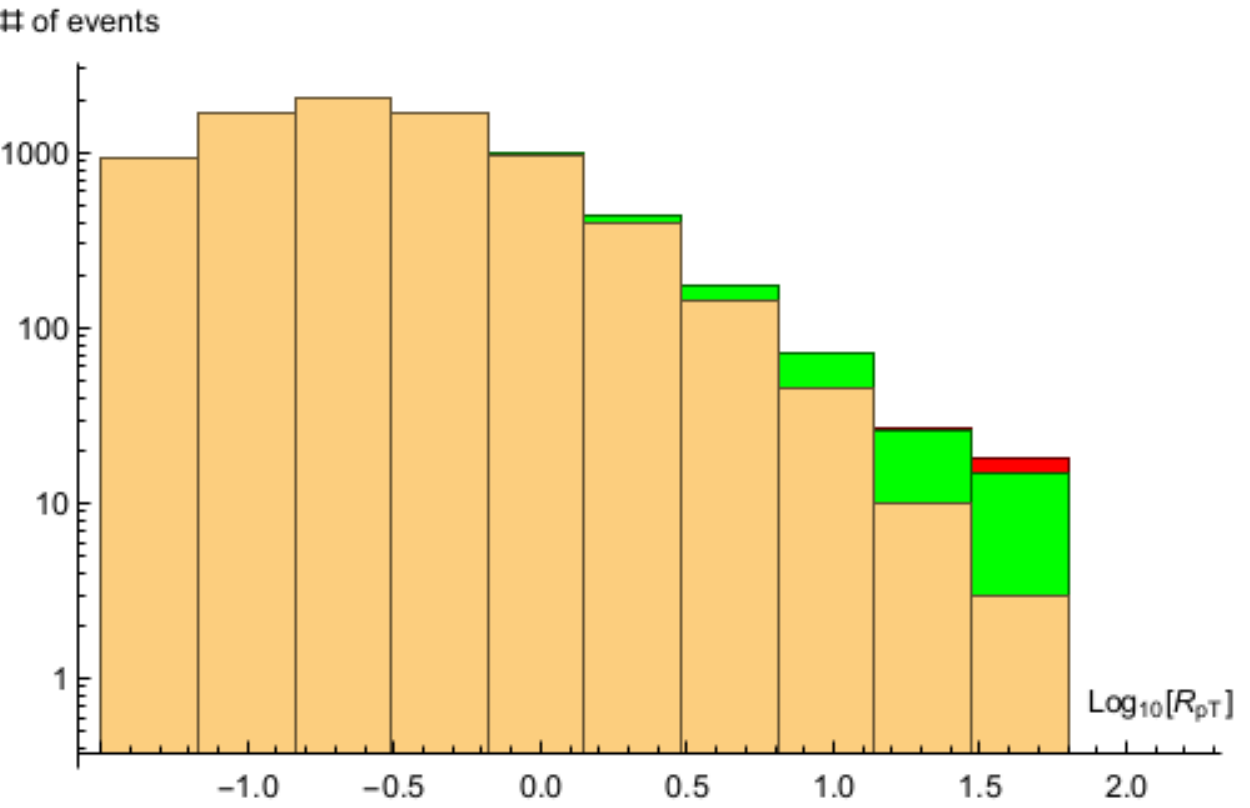} 
}
\end{center}
\caption{\sl Like in Fig.~\ref{fig:distrosIndividualCWgeq} for $f_{\varphi\Box}=-3.4$ TeV$^{-2}$.}
\label{fig:plotHisto19SMcphiBox_m3_4}
\end{figure}
\begin{figure}[h]
\begin{center}
\mbox{
\includegraphics[scale=0.6]{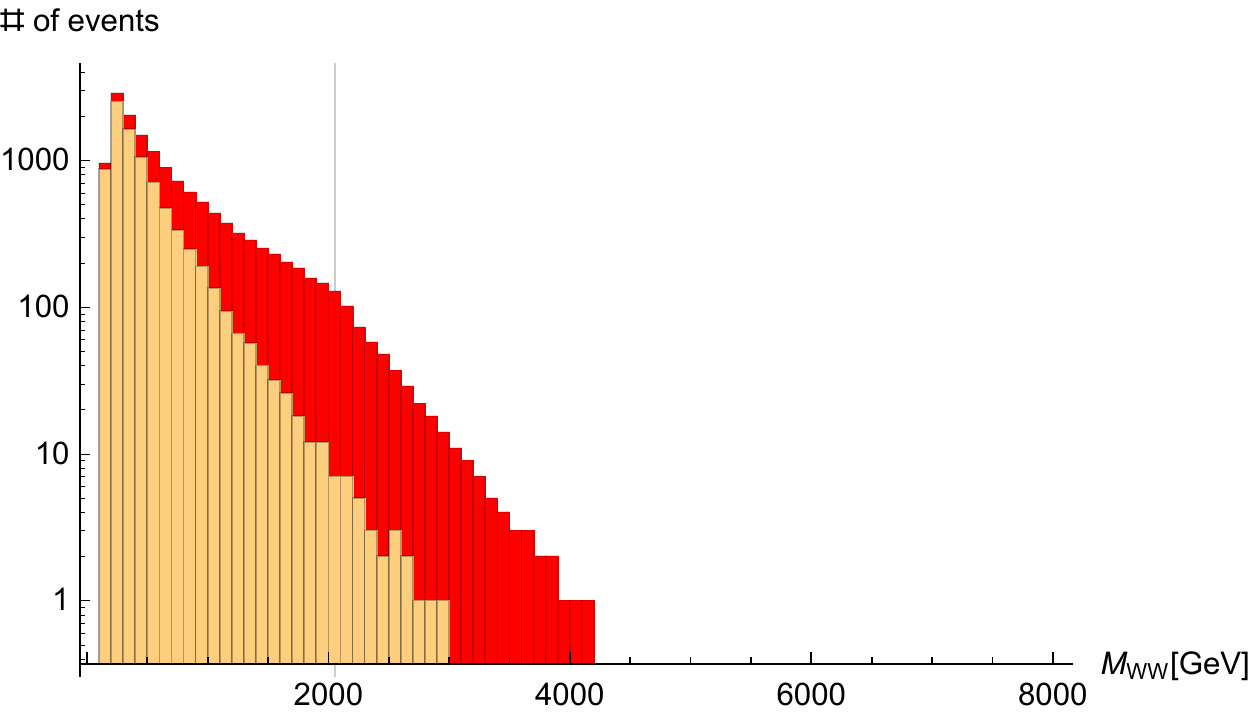} 
\includegraphics[scale=0.6]{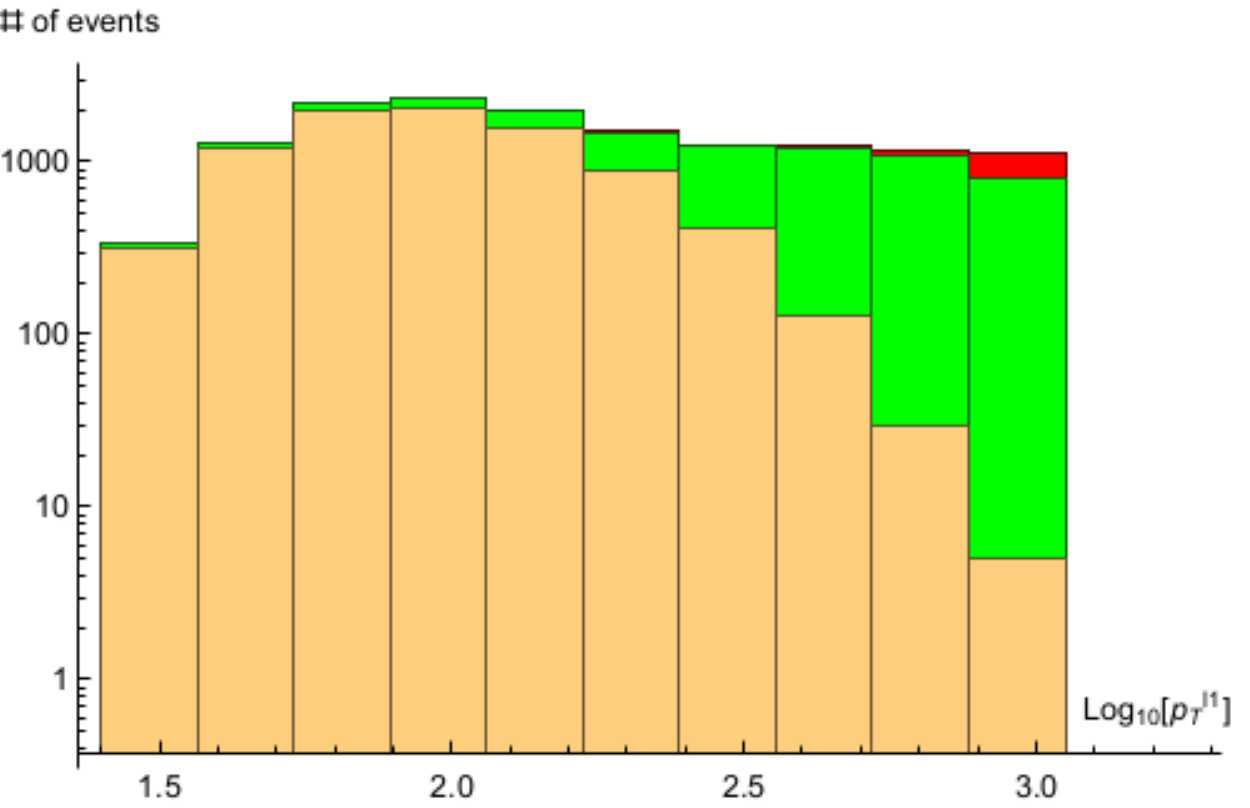} 
}
\end{center}
\caption{\sl Like in Fig.~\ref{fig:distrosIndividualCWgeq} for  $f_{W}=+1.$ TeV$^{-2}$.}
\label{fig:plotHisto19SMcW_1}
\end{figure}

The distributions in $M_{WW}$ together with the most sensitive variable, are shown in Figs.~\ref{fig:plotHisto19SMcphiW_1_6},~\ref{fig:plotHisto19SMcphiD_m2_7},~\ref{fig:plotHisto19SMcphiBox_m3_4} and~\ref{fig:plotHisto19SMcW_1} for operators  $Q_{\varphi D}$, $Q_{\varphi W}$, 
$Q_{\varphi\Box}$ and $Q_{W}$, respectively (for $f_{i=\varphi D,\varphi W,\varphi\Box}$ at the relevant, either positive or negative, boundary).

As above, it is worth noticing that for all  operators, the bulk of the SM EFT effect is within the EFT-valid region. In terms of standard deviations ($\sigma$), the maximal discrepancies for $Q_{\varphi D}$, $Q_{\varphi W}$, $Q_{\varphi\Box}$ at the HL-LHC allowed by current data read as $4.4,15.$, $12.\,\sigma$, respectively. Hence these operators cannot be, in general, simply neglected in studies for HL-LHC prospects of discovery potential via the EFT approach. 

The possible effects from $Q_W$ are, in accord with Sec.~\ref{sec:W}, exceptionally large -- the obtained discrepancy at the HL-LHC is $O(100.)\sigma$. 
In this case the exact number of $\sigma$s may vary depending on the chosen method of unitarization, as anticipated in Sec.~\ref{subsec:unitarityBounds}. For the scope of this paper it is rather inessential, so we report only the order of magnitude.  Clearly, signal significance can be huge, much above $5\sigma$ even in the ``EFT-controlled'' estimate, therefore not at all negligible in VBS data analyses in the HL-LHC perspective.
 Moreover, a simple rescaling to current luminosity, would imply large possible discrepancies already in the currently collected LHC dataset. Though counter-intuitive at first sight, it is partially a result of truncating the cross-section consistently at dimension-6, which, as already discussed in the introduction, leads to conservative bounds from di-boson production channel. Although the same-sign $WW$ scattering studies illustrate the necessity for better theoretical understanding (and as a consequence improvement) in setting constraints based on di-boson production at LHC, it also shows (i) large potential in VBS processes to set constraints on the $f_W$ coefficient or alternatively (ii), by comparison of the red and green histograms in the right plots in Fig.~\ref{fig:distrosIndividualCWgeq} and~\ref{fig:plotHisto19SMcW_1}, large discovery potential which is quantifiable by (seizable) ``EFT-triangle''~\cite{Kalinowski:2018oxd}; interestingly, the EFT-triangles for dimension-8 interactions are very limited~\cite{Kalinowski:2018oxd,Chaudhary:2019aim}, due to EFT-validity issues caused by necessarily large high-$M_{WW}$ tails, which makes indirect searches in transversal $WW$ scattering via dimension-6 even more attractive.

\section{Epilogue}
\label{sec:conclusions}

In this article, we performed a detailed study of SM EFT effects from
 dimension-6 operators
in a VBS process with like-sign-$W$ production proceeding to leptonic final states. 
We studied all relevant  dimension-6 operators at tree level which are
 responsible for NP in VBS including 
``background" operators responsible for NP in non-VBS parts of the full process
$pp\rightarrow jj\ell\ell'\nu\nu'$.  In addition, in our analysis 
we exploited all available constraints 
on SM EFT 
Wilson coefficients and also derived the relevant perturbative unitarity bounds. The latter are applied to obtain quantitative estimate on the role of high-energy tails that cannot be described within the EFT approach.
 We have presented useful analytical formulae which illustrate and support
the robustness of our simulations' results.

\begin{table}[h!]%
\begin{center}
\begin{tabular}{|c||c|c|c|c|} \hline
     & $\mathcal{O}_{W}$ & $\mathcal{O}_{\varphi\Box}$ & $\mathcal{O}_{\varphi D}$ & $\mathcal{O}_{\varphi W}$ \\ \hline
 $\sigma^{HL-LHC}\leq$ & O(100.) & 12. & 4.4 & 15.\\ \hline
\end{tabular}
\end{center}
\caption{\sl Maximal effect estimates in standard deviations ($\sigma$)  for the relevant dimension-6 interactions in same-sign $WW$ scattering allowed within present constraints as reported in the global fit analysis of~\cite{Dawson:2020oco} all normalized to HL-LHC.}
\label{tab:sigmasGlobal}
\end{table}

We found that the role of dimension-6 operators in modifying the $W^+W^+\rightarrow W^+W^+$ process can \emph{not} be neglected in full generality, particularly for perspective studies at the HL-LHC  based on current constraints. This is particularly true when the constraints on dimension-6 operators are taken from the global fit type analyses. The potential effect based on current global fit limits is summed up in Table~\ref{tab:sigmasGlobal} for the relevant operators. 
It shows a particular sensitivity to strong dynamics for transverse $W$-polarization, effects of which can be conservatively estimated with dimension-6 truncated amplitudes.
Plausibly, the background operators are already constrained strongly enough to claim they play no role in $WW$-scattering process even under the HL-LHC perspective, with the  exception of $Q_{\varphi q}^{(3)}$ operator (and potentially $Q_{\varphi ud}$ and dipole operators).  These VBS-polluting
operators, however, feature very distinctive dynamics, and they can be separated by studying different kinematic distributions.

\section*{Acknowledgements}
We would like to thank Janusz Rosiek for his help in handling
 input parameter schemes in SmeftFR code. AD would
like to thank Lambros Trifyllis and Kristaq Suxho for cross-checking a number of  helicity amplitudes. 
PK is supported by the Spanish MINECO project FPA2016-78220-C3-1- P (Fondos FEDER),  A-FQM-211-UGR-18 (Junta de Andalucía, Fondos FEDER)
 and by National Science Centre, Poland, the PRELUDIUM project under contract 2018/29/N/ST2/01153. 
The research of MS has been partly supported by VBSCan (COST action CA16108).
We acknowledge partial support from National
Science Centre, Poland, grant\\ \mbox{DEC-2018/31/B/ST2/02283}.


\begin{appendices}

\section{Helicity Amplitudes and Cross-Sections}
\label{app:A}
Probably the best way to write down the anatomy of a cross section, especially for vector boson scattering, is to calculate first the helicity amplitudes for the given
process. Being definitive, lets assume the elastic scattering of $W^+$-gauge bosons,
\begin{equation}
W^+(p_1,\lambda_1) + W^+(p_2,\lambda_2) \to W^+(p_3,\lambda_3) +  W^+(p_4,\lambda_4)\;. \label{eq:wpwmwpwm}
\end{equation} 
We consider the scattering plane to be the $xz$ plane with $\vec{p}_1$ in $+\hat{z}$
direction, with the outgoing particle-3 momentum $p_{3x}$ in positive $\hat{x}$-direction. More specifically,
the kinematics are given by
\begin{eqnarray}
p_1 &=& (E,0,0,p)\;, \\
p_2 &=& (E,0,0,-p)\;, \\
p_3 &=& (E, p \sin\theta, 0, p \cos\theta ) \;, \\ 
p_4 &=& (E, -p \sin\theta, 0, -p \cos\theta ) \;.
\end{eqnarray}
The center-of-mass 
scattering angle $\theta$ is restricted in $\theta \in [0,\pi]$ interval. For process
\eqref{eq:wpwmwpwm} we have 81 helicity amplitudes. However, not all of them 
are distinct because of $C-$, $P-$, and $T-$ transformations as well as rotational invariance of the $S$-matrix. 
For example, CPT and CP transformations read as
\begin{eqnarray}
\mathrm{CPT-symmetry}\; : 
\qquad \mathcal{M}_{\lambda_1, \lambda_2, \lambda_3 , \lambda_4}(\theta) &=&
\mathcal{M}_{-\lambda_3, -\lambda_4, -\lambda_1 , -\lambda_2} (\theta)\;, \\
\mathrm{CP-symmetry}\; : \qquad \mathcal{M}_{\lambda_1, \lambda_2, \lambda_3 , \lambda_4} (\theta) &=&
\mathcal{M}_{-\lambda_2, -\lambda_1, -\lambda_4 , -\lambda_3} (-\theta)\;,
\end{eqnarray}
relate the amplitudes up-to a phase factor~\cite{Denner:1997kq,deRham:2017zjm}.

For an elastic polarized cross section like $W^+(\lambda_1) W^+(\lambda_2) \to W^+(\lambda_3) W^+(\lambda_4)$ we have 
\begin{equation}
\biggl ( \frac{d \sigma}{d \Omega} \biggr )_{\lambda_1 \lambda_2 \lambda_3 \lambda_4} = \frac{1}{64 \pi^2 s} \: |\mathcal{M}_{\lambda_1 \lambda_2 \lambda_3 \lambda_4}(s,\cos\theta) |^2 \;,
\end{equation} 
while for unpolarized ones we have to sum over final and initial helicities and average over initial helicities 
\begin{equation}
\biggl ( \frac{d \sigma}{d \Omega} \biggr )_{\mathrm{unpolarized}} = \frac{1}{9} \sum_{\lambda_1 , \lambda_2 = -1}^1  \sum_{\lambda_3 , \lambda_4 = -1}^1  \:
\biggl ( \frac{d \sigma}{d \Omega} \biggr )_{\lambda_1 \lambda_2 \lambda_3 \lambda_4}\;.
\end{equation}
The integrated cross-section is defined as 
\begin{equation}
\sigma(s) = \int_0^{2\pi} d\varphi \int_{\theta_{cut}}^{\pi -\theta_{cut}} d\theta \sin\theta \: \frac{d\sigma}{d\Omega}\;,
\end{equation}
where $\theta_{cut}$ is an angular cut in scattering angle.

We append in \ref{app:1} the helicity  amplitudes $ \mathcal{M}_{\lambda_1 \lambda_2 \lambda_3 \lambda_4} $ in SM EFT with dimension-6 operators by keeping the 
SM contributions  to $O(1)$ and up to  leading $O(s)$-amplitudes associated with the  operators appeared in Table~\ref{tab:ops}.  We also append in \ref{app:2} 
a relevant to our discussion, 
partial list of amplitudes for SM EFT dimension-8 operators at leading $O(s^2)$.
In calculating the amplitudes,
we follow the notation of ref.~\cite{dedes:2017zog} for vertices  while we use the  {\tt SmeftFR}~\cite{Dedes:2019uzs} to output Feynman Rules as 
input to  {\tt FeynArts}~\cite{Hahn:2000kx}  and {\tt FormCalc}~\cite{Hahn:1998yk} for amplitude calculations. In case of longitudinal vector boson scattering we have checked our results analytically with the Goldstone boson equivalence theorem~\cite{Cornwall:1974km,Vayonakis:1976vz,Lee:1977eg,Chanowitz:1985hj,Gounaris:1986cr,Yao:1988aj,Bagger:1989fc}.

\subsection{Helicity amplitudes for $W^+(\lambda_1) W^+(\lambda_2) \to W^+(\lambda_3) W^+(\lambda_4)$ in SM EFT}
\label{app:1}

\noindent The SM leading $\mathcal{O}(1)$ contributions, where $\lambda$ is the Higgs-self coupling and 
$\bar{g}$ the SM EFT improved  $SU(2)_L$-gauge coupling, are:
\begin{align}
\mathcal{M}_{\pm\pm\pm\pm} &= -\frac{8 \bar{g}^2}{1-\cos^2\theta}\;,
\label{eq:M++++} \\
\mathcal{M}_{\pm\mp\mp\pm} &= -2 \bar{g}^2 \frac{(1-\cos\theta )}{(1+\cos\theta)} \;, \\
\mathcal{M}_{\pm\mp\pm\mp} &= -2 \bar{g}^2 \frac{(1+\cos\theta)}{(1-\cos\theta)}\;, \\
\mathcal{M}_{0\pm 0 \pm} &=  \mathcal{M}_{\pm 0  \pm 0} =  -2  \bar{g}^2  \biggl (\frac{1}{1-\cos\theta}\biggr ) \;,\\
\mathcal{M}_{\pm 0 0 \pm} &=  \mathcal{M}_{ 0  \pm \pm 0} =  -2 \bar{g}^2  \biggl (\frac{1}{1+\cos\theta}\biggr ) \;,\\
\mathcal{M}_{0 0 0 0} &= \frac{1}{2} (\bar{g}^2 + \bar{g}^{\prime 2} ) \biggl (1- \frac{4}{\sin^2\theta}\biggr ) - 2 \lambda  \;.
\label{amp:SMpppp}
\end{align}
The SM EFT leading-$s$ contributions from CP-Conserving couplings associated to
operators in the CPC row in Table~\ref{tab:ops} are:
\begin{align}
\mathcal{M}_{\pm\mp\mp\mp} &=\mathcal{M}_{\mp\pm\mp\mp} = \mathcal{M}_{\mp\mp\pm\mp} = \mathcal{M}_{\mp\mp\mp\pm} = - 6\, \bar{g} C^W  \: \frac{s}{\Lambda^2} \;,\label{amp:ppppTTTT} \\
\mathcal{M}_{\pm\pm\mp\mp} &= 12\, \bar{g} C^W \:  \frac{s}{\Lambda^2} \;, \\
\mathcal{M}_{0\pm 0\mp} &= \mathcal{M}_{\pm 0 \mp 0} = -\frac{3}{4}\, \bar{g} C^W (3 + \cos\theta) \: \frac{s}{\Lambda^2}
+ C^{\varphi W} \: (1-\cos\theta) \: \frac{s}{\Lambda^2}\;, \\
\mathcal{M}_{0\pm\mp0} &=  \mathcal{M}_{\pm00\mp}= \frac{3}{4}\, \bar{g} C^W (3-\cos\theta) \: \frac{s}{\Lambda^2} - C^{\varphi W} \: (1+\cos\theta) \: \frac{s}{\Lambda^2}\:  \;,\\ 
\mathcal{M}_{0000} &=  \biggl (2 C^{\varphi\Box} +  C^{\varphi D} \biggr ) \: \frac{s}{\Lambda^2}\;. 
\label{amp:ppppLLLL}
\end{align}

\noindent The SM EFT leading-$s$ contributions from CP-Violating couplings associated to
operators in the CPV row in Table~\ref{tab:ops} are:
\begin{align}
\mathcal{M}_{++--} &= -\mathcal{M}_{--++} = -12 \, \bar{g} C^{\widetilde{W}}  \: \frac{s}{\Lambda^2} \;, \\[2mm]
\mathcal{M}_{+---} &=  \mathcal{M}_{-+--} = - \mathcal{M}_{--+-} = - \mathcal{M}_{---+} = 6 \, \bar{g} C^{\widetilde{W}} \: \frac{s}{\Lambda^2} \nonumber \\
&= -\mathcal{M}_{-+++} = -\mathcal{M}_{+-++} = \mathcal{M}_{++-+} = \mathcal{M}_{+++-}  \;, \\[2mm]
\mathcal{M}_{0+-0} &=  \mathcal{M}_{+00-} =  - \frac{3}{4} \bar{g} C^{\widetilde{W}} (3- \cos\theta )  \: \frac{s}{\Lambda^2} 
+ C^{\varphi \widetilde{W}} (1+ \cos\theta )  \: \frac{s}{\Lambda^2} = \nonumber \\
&= -\mathcal{M}_{0-+0}  =- \mathcal{M}_{-00+} \;, \\[2mm]
\mathcal{M}_{0+0-} &=  \mathcal{M}_{+0-0} =   \frac{3}{4} \bar{g} C^{\widetilde{W}} (3+ \cos\theta )  \: \frac{s}{\Lambda^2} 
- C^{\varphi \widetilde{W}} (1- \cos\theta )  \: \frac{s}{\Lambda^2} = \nonumber  \\
&= - \mathcal{M}_{0-0+} =- \mathcal{M}_{-0+0} \;. 
\label{amp:ppppCPV}
\end{align}

The SM results for the amplitudes agree with refs.~\cite{Gounaris:1993aj,Vayonakis:1976vz}.
We convert the set of parameters $\{\bar{g},\bar{g}',\bar{v},\lambda \}$, to a  set of input observables\footnote{See for example ref.~\cite{Dedes:2019bew} for the conversion.} $\{ G_F,m_W,m_Z,m_h\}$  and write out analytically the  cross-sections in the c.m. frame for Longitudinal and Transverse gauge bosons averaging  over the initial helicities. We find at leading order in $s$:
\begin{align}
\sigma_{TTTT}(s) &= \frac{8}{\pi s} (G_F m_W^2)^2 c \left ( \frac{9-c^2}{1-c^2} \right ) + \frac{36 \sqrt{2} c}{\pi s} (G_F m_W^2) \left (|C^W|^2 + |C^{\widetilde{W}}|^2 \right ) \left (\frac{s}{\Lambda^2} \right )^2\;, \label{eq:WpTTTT}
\\[3mm]
\sigma_{LLLL}(s) &= \frac{1}{2\pi s} (G_F m_Z^2)^2 \left [ c \frac{9-c^2}{1-c^2} - 2 (c + 2 L_c) \frac{m_h^2}{m_Z^2} + \left (\frac{m_h^2}{m_Z^2} \right )^2 \right ] \nonumber \\[2mm]
&+ \frac{\sqrt{2}}{4\pi s} (G_F m_Z^2) (2 C^{\varphi\Box} + C^{\varphi D}) \left ( \frac{s}{\Lambda^2} \right ) \left [ (c + 2 L_c) - c \frac{m_h^2}{m_Z^2} \right ] \nonumber \\[2mm]
&+ \frac{c}{16 \pi s} \left (\frac{s}{\Lambda^2} \right )^2 \left (2 C^{\varphi\Box} + C^{\varphi D} \right )^2 \;, \label{eq:WpLLLL} \\[3mm]
\sigma_{LTLT}(s) &= \sigma_{TLTL}(s) = \sigma_{TLLT}(s) = \sigma_{LTTL}(s) = \frac{8}{\pi s} (G_F m_W)^2 \left (\frac{c}{1-c^2} \right ) \nonumber \\[2mm]
&- \frac{2^{1/4}}{16\pi s} (G_F m_W^2)^{1/2} \left (\frac{s}{\Lambda^2} \right )^2
c (9-c^2) \: \Re e \left ( C^W C^{\varphi W*} + C^{\widetilde{W}} C^{\varphi \widetilde{W}*} \right ) \nonumber \\[2mm]
&+ \frac{3}{16 \pi s} \left ( \frac{s}{\Lambda^2} \right )^2 \left [
\frac{c(3+c^2)}{9} \left (|C^{\varphi W}|^2 + |C^{\varphi \widetilde{W}}|^2 \right ) +
\frac{ \sqrt{2} c (27 + c^2)}{4} (G_F m_W^2) \left (|C^W|^2 + |C^{\widetilde{W}}|^2 \right ) \right ]\;. \label{eq:WpLTLT}
\end{align}
Since $\cos\theta\to 0$ is not attainable we have used a cut at small angle $\theta_{cut}$
and integrate over $\theta_{cut}$ to $\pi -\theta_{cut}$.
In eqs.~\eqref{eq:WpTTTT}-\eqref{eq:WpLTLT}   
we used the abbreviations: $c\equiv \cos\theta_{cut}$, and 
$L_c \equiv \log \left (\frac{1-c}{1+c} \right )$. Numerically, if $\theta_{cut}=\pi/18$
then $(1-c^2)^{-1} \approx 33$ and $L_c \approx -5$ and therefore, cross-sections are 
dominated by angles close to the beam.

\subsection{Helicity amplitudes for pure transversal $W$'s with up to $d=8$ operators}
\label{app:2}
One of the important outcomes of our analysis is that VBS cross-section 
is strongly affected  by the operator $O_W= \epsilon^{IJK} W_{\mu\rho}^I W^{\rho\sigma , J} W_\sigma^{\mu , K}$ which enters in the transverse-$W$ four-point interactions. 
Since this operator is not affected by  Higgs field redefinitions after EW symmetry breaking, the
inclusion of pure transverse $W^4$ operators is straightforward.
For pure interactions $W^4$ the  basis  for dimension-8 operators  
reads~\cite{Remmen:2019cyz,Murphy:2020rsh,Li:2020gnx,Yamashita:2020gtt}\footnote{Our notation for dimension-8 operators follows closely the Warsaw basis 
article~\cite{Grzadkowski:2010es} where operators contain only gauge fields and not weighted by gauge couplings. 
For comparisons, our $C^{t_i}$ coefficients are $g^4/4$ times  those of Ref.~\cite{Yamashita:2020gtt}.}
\begin{align}
O_{t0} \ &= \ (W_{\mu\nu}^I W^{\mu\nu , I} ) (W_{\alpha \beta}^J W^{\alpha\beta , J} )\;, \label{Ot0}\\[2mm]
O_{t1} \ &= \ (W_{\alpha\nu}^I W^{\mu\beta , I} ) (W_{\mu \beta}^J W^{\alpha\nu , J} )\;, \\[2mm]
O_{t2} \ &= \ (W_{\alpha\mu}^I W^{\mu\beta , I} ) (W_{\beta \nu}^J W^{\nu\alpha , J} )\;, \\[2mm]
O_{t10} \ &= \ (W_{\mu\nu}^I \widetilde{W}^{\mu\nu , I} ) (W_{\alpha \beta}^J \widetilde{W}^{\alpha\beta , J} )\;.\label{Ot10}
\end{align}
The \textit{leading-$s$}, CP-even, helicity amplitudes up to $1/\Lambda^4$, is symbolically written as,
\begin{equation}
\mathcal{M} \ = \ \mathrm{SM} + \mathrm{SM} \cdot \mathrm{dim6}   + (\mathrm{dim6})^2 +
\mathrm{SM} \cdot \mathrm{dim8} \;,
\end{equation}
are found to be 
\begin{align}
\mathcal{M}_{\pm\pm\pm\pm} \ &= \ -\frac{8 \bar{g}^2}{1-\cos^2\theta} \ + \ 
4 \: (2\, C^{\: t1}+C^{\: t2}) \frac{s^2}{\Lambda^4}\;,\label{eq:d81}\\[2mm]
\mathcal{M}_{\pm\mp\pm\mp} \ &= \ -2 \bar{g}^2 \frac{1+\cos\theta}{1-\cos\theta} 
\ -\ \frac{9}{2}\: |C^W|^2 \cos^2 \left ( \frac{\theta}{2} \right ) (3-\cos\theta) \frac{s^2}{\Lambda^4} \nonumber \\
\ &+ \ (\cos\theta + 1)^2\: ( 2\, C^{\: t0} + C^{\: t1} + C^{\: t2} - 2\, C^{\: t10})
\frac{s^2}{\Lambda^4} \;, \\[2mm]
\mathcal{M}_{\pm\pm\mp\mp} \ &= 12 \,\bar{g} C^W \frac{s}{\Lambda^2}  \ - \
\frac{9}{2} \: (3 -\cos^2\theta) \: |C^W|^2 \frac{s^2}{\Lambda^4} \nonumber \\
\ &+ \ (1+\cos^2\theta)\: ( 4\: C^{\, t0} + 2\, C^{\, t1} + C^{\, t2} + 4\, C^{\, t10} ) \frac{s^2}{\Lambda^4} + 8\, C^{\, t1} \frac{s^2}{\Lambda^4} \;, \label{eq:dim62} \\[2mm]
\mathcal{M}_{\pm\mp\mp\mp} \ &= -6\, \bar{g} C^W \frac{s}{\Lambda^2} \;.
\label{eq:d84}
\end{align}
The multiplicities of the above helicity amplitudes into the transversely polarized
 cross-section $\sigma_{TTTT}$, are 
2:4:2:8, respectively. All other contributions not written explicitly in eqs.~\eqref{eq:d81}-\eqref{eq:d84}, growing at most like 
 $(s\, v^2/\Lambda^4)$, are neglected. It is amusing to note that 
 $(\mathrm{dim6})^2$ terms do not involve gauge couplings (as they should) 
 in broken phase:
 they result from the sum of $Z$ and $\gamma$ tree-diagrams with identical Lorentz structures (see $WWZ$ and $WW\gamma$ vertices in \cite{dedes:2017zog}). On the other hand, dimension-8 
 leading-$s$ contributions arise from contact terms only.

\end{appendices}

\bibliography{VBS-SMEFT}{}
\bibliographystyle{JHEP}

\end{document}